\def\fullversion{1}

\ifnum\fullversion=1
\documentclass[11pt]{article}
\usepackage[utf8]{inputenc} 
\usepackage[T1]{fontenc}    
\usepackage{hyperref}       
\usepackage{url}            
\usepackage{booktabs}       
\usepackage{amsfonts}       
\usepackage{nicefrac}       
\usepackage{microtype}      

\usepackage{amstext}
\usepackage{amsmath}
\usepackage{amsthm}
\usepackage{amssymb}
\usepackage{bm}
\usepackage{wrapfig}
\usepackage{amsfonts, amsmath}
\usepackage[utf8]{inputenc}
\usepackage{graphicx}
\usepackage{bbm, comment}
\usepackage{subfigure}
\usepackage{color}
\usepackage{float}
\usepackage{wrapfig}
\usepackage{url}

\usepackage{MnSymbol,wasysym,marvosym,ifsym}

\usepackage{tabulary}
\usepackage{booktabs}
\usepackage{diagbox}

\usepackage[top=1.in, bottom=1.in, left=1.in, right=1.in]{geometry}
\usepackage[numbers]{natbib}

\newtheorem*{theorem*}{Theorem}
\newtheorem{theorem}{Theorem}[section]
\newtheorem{lemma}[theorem]{Lemma}
\newtheorem{fact}[theorem]{Fact}
\newtheorem{claim}[theorem]{Claim}
\newtheorem{assumption}[theorem]{Assumption}

\newtheorem{example}[theorem]{Example}

\newtheorem{definition}[theorem]{Definition}

\else

\documentclass[format=acmsmall, review=true]{acmart}

\usepackage{acm-ec-18}

\usepackage{booktabs}

\usepackage{diagbox}

\usepackage{bbm,bm}

\usepackage{wasysym}

\fi

\usepackage{algorithm}
\usepackage[noend]{algpseudocode}

\makeatletter
\def\BState{\State\hskip-\ALG@thistlm}
\makeatother

\usepackage[most]{tcolorbox}

\makeatletter
\newtcolorbox{note}[1][]{%
 breakable,
  enhanced jigsaw, 
  borderline west={1pt}{0pt}{black}, 
  sharp corners, 
  boxrule=0pt, 
  fonttitle={\large\bfseries},
  coltitle={black},  
  title={ },  
  attach title to upper, 
  right=0pt,
  top=5pt,
  bottom=5pt,
  frame hidden,
  baseline={\tcb@height-2\kvtcb@boxsep+\baselineskip-2\lineskip}, 
  #1
}
\makeatother

\usepackage{blindtext}

\newcommand{\E}{\mathbb{E}}

\DeclareMathOperator*{\argmax}{arg\,max}

\newcommand{\faithful}{potent }
\newcommand{\Faithful}{Potent }

\newcommand\numberthis{\addtocounter{equation}{1}\tag{\theequation}} 

\newcommand{\methodset}{M}

\usepackage{xcolor}

\newcommand\gs[1]{{}}
\newcommand\yk[1]{{}}

\allowdisplaybreaks

\begin{document}

\title{Eliciting Expertise without Verification}
\author{Yuqing Kong\\ University of Michigan \and Grant Schoenebeck\\ University of Michigan}
\date{}

\maketitle

\begin{abstract}

A central question \footnote{This work is supported by the National Science Foundation, under grant CAREER\#1452915, CCF\#1618187 and AitF\#1535912.} of crowdsourcing is how to elicit expertise from agents. This is even more difficult when answers cannot be directly verified.  A key challenge is that sophisticated agents may strategically withhold effort or information when they believe their payoff will be based upon comparison with other agents whose reports will likely omit this information due to lack of effort or expertise.       

Our work defines a natural model for this setting based on the assumption that \emph{more sophisticated agents know the beliefs of less sophisticated agents}. 

We then provide a mechanism design framework for this setting.  From this framework, we design several novel mechanisms, for both the single and multiple tasks settings, that (1) encourage agents to invest effort and provide their information honestly; (2) output a correct ``hierarchy'' of the information when agents are rational.  

\end{abstract}

\section{Introduction}

Crowdsourcing, outsourcing tasks to a crowd of workers (e.g. Amazon Mechanical Turk, peer grading for massive open online courses, scholarly peer review, and Yahoo answers), is a fast, cheap, and effective method for performing simple tasks even at large scales.  To attract a large number of workers, crowdsourcing is usually open to the public rather than just professional experts.  This makes crowdsourcing a challenge for more complicated tasks where agents with different levels of expertise may honestly report different answers. For example, in peer grading, some students may be incapable of accurate grading, 
missing errors other students point out.  

One solution is to spot-check random tasks or equivalently to insert ``gold-standard'' tasks to which the mechanism already knows the answer.  However, such techniques are expensive and even impossible for tasks where the agents' opinions can be very subjective (e.g. peer grading essay assignments) or individualized (e.g. experience at a restaurant).  

\subsection{Our setting}
This work is about the ``peer-prediction'' setting where such verification is unavailable or undesirable.  Already, a large line of work in the peer prediction (PP) area \citep{2016arXiv160501021K,witkowski2014robust,radanovic2015incentive, 2016arXiv160307319K, 2016arXiv160303151S, KG16, radanovic2015incentives, witkowski2013learning, witkowski2012peer,kamble2015truth,dasgupta2013crowdsourced,radanovic2014incentives,riley2014minimum,zhang2014elicitability,faltings2014incentives,radanovic2013robust,jurcafaltings06,jurcafaltings09,jurcafaltings07,jurcafb08,jurca2011incentives,lambert2008truthful,goelrp09,MRZ05,papakonstantinou2011mechanism,chen2012predicting,prelec2004bayesian,liu2016learning,liu2016sequential} designed mechanisms for agents with similar expertise (without verification).

These papers study various settings.  In the \emph{single-task} setting~\cite{MRZ05,prelec2004bayesian} agents are asked one multiple-choice question but typically share a common prior over the responses; in the \emph{multi-tasks} setting~\cite{dasgupta2013crowdsourced,2016arXiv160501021K,2016arXiv160303151S}  agents are assigned a batch of a priori similar multiple-choice questions.  We consider both of these settings.

In the \emph{detail-free} setting, the mechanism is not required to know information about the (common) prior distribution.  Our mechanisms are generally in this setting, but some of them do require information about the different levels of expertise/effort agents can exhibit. 


However, the aforementioned peer-prediction papers do not consider settings where \begin{enumerate} \item  \textbf{Agents have different levels of expertise} or \item \textbf{A lack of effort can systemically bias agents' reports}.  \end{enumerate}

The following two tasks exemplify settings 1) and 2) respectively:

\begin{example}\label{eg:state}
Which state (from a list of all 50 states) in the United States of America is closest to Africa? (Single-task)
\end{example}

\begin{example}\label{eg:peer}
Peer grading several essays by providing a grade from the set $\{1, 2, 3, 4, 5\}$. (Multi-tasks)
\end{example}

In the first example, an agent can guess randomly (no effort), look up the correct answer (full effort), or guess at the correct answer (partial effort).  Most people will guess Florida, even though experts will know the correct answer is Maine. Thus differing levels of expertise yield different answers.  

In the second example, a student can, instead of carefully grading (full effort) or assigning a random grade (zero effort), quickly check the name of the top of the paper and spot check the grammar (partial effort).  Thus partial effort can systematically bias agents: consider an essay from a top student in an impeccable pose, but which contains large conceptual errors.  Here partial effort can give some information about the correct answer, but also enable agents to ``coordinate'' on an incorrect answer.

Gao, Wright, and Leyton-Brown~\cite{2016arXiv160607042G} show that the effects of the settings 1) and 2) are devastating to  previous peer-prediction mechanisms, which generally fail in motivating the agents to invest effort for ``expensive signals'' when ``cheap signals'' (that ensure agreement and may even be correlated with the sought signal) exist.  The main (very high-level) idea behind previous peer-prediction mechanisms can be understood as a ``clever majority vote'', every agent is paid according to a specific similarity between her and her peer.  Thus, they point out that in the peer-grading example, coordinating on just checking the grammar can guarantee good agreement with other agents, but with substantially reduced effort.  

In fact, Gao et al point out that things are likely even worse than this.  If the cheap signals correlate more than the expensive signals, then the peer-prediction techniques incentivize agents to not report the true answer, but instead focus on cheap signals!  For example, in the essay grading above, it is likely that assessments of grammatical correctness will agree more than assessments of overall essay quality.  Because of this, peer-prediction mechanisms will pay agents more overall for lower-quality information.  In Example~\ref{eg:state}, even if agents know the answer is Maine, they may report Florida, expecting that most others will do likewise.

\gs{More DIRE!}

Such behavior undermines the goal of applying crowdsourcing to increasingly complex tasks, and, in fact, undercuts any application of crowdsourcing to perform any task where the answers are not ``common knowledge."  The field must overcome this key challenge of rewarding rather than suppressing expertise in order to begin the project of expanding crowdsourcing beyond simple labeling tasks.

\subsection{Key Insight and Assumption}

Previous peer-prediction mechanisms treat all information (cheap/expensive) equally, and thus agents lack an incentive to invest the effort to obtain expensive signals. Moreover, even when an expert can easily obtain the expensive signal, the previous mechanism discourage her from providing it when she believes the non-experts will disagree.

A successful mechanism must break the symmetry between weak and expensive signals and between expert and non-expert signals.  We propose the following natural assumption which will allow a mechanism to break this symmetry.

\begin{assumption}\label{assume:hier}
Agents with high effort or expertise know the beliefs of agents with less effort or less expertise.   
\end{assumption}

We can see that this assumption is very natural in Example~\ref{eg:state} and Example~\ref{eg:peer}.  Agents who look up or know the answer in  Example~\ref{eg:state} also know most people will answer ``Florida.''  Agents that carefully grade an essay can also approximate the score of an agent who spends very little effort. We will define a hierarchical information structure (Section~\ref{sec:framework}) to naturally capture Assumption~\ref{assume:hier}.

Our mechanisms solicit not only agents'  own opinions but also their predictions for the opinions of the other agents who have less information. This differs with the previous peer prediction mechanisms which ask agents to provide their predictions for \emph{all} other agents' opinions.
For example, in the peer-grading example, we might ask agents to report their own evaluation, and to \emph{optionally} report one or more low-effort / low expertise evaluations (e.g. scores based on grammar, thesis statement, student name, naive reading, etc).

\subsection{Our contribution}  
\begin{description}
  \item[Modeling:] We define a hierarchical information structure (the higher level information is more valuable and harder to obtain)  model  that captures the hierarchical relationship between the different expertise;
  \item[Mechanism design:] We employ information theory tools to provide a mechanism design framework for the aforementioned hierarchical setting---Hierarchical Mutual Information Paradigm (HMIP)---which 1) pays the higher level information more and encourages agents to invest effort for different levels of information according to their own abilities 2) incentivizes agents to report honestly; and 3) only pays agents for information they actually learn. We apply our mechanism design framework to create the following mechanisms:
  
  \begin{description}
     \item[Multi-HMIM] which works in the multi-task setting even for a small number of tasks but requires the mechanism to know the hierarchical information structure.
     \item[Learning-based Multi-HMIM] which works in the multi-task setting even when the mechanism does not know the hierarchical information structure; however requires a large number of tasks.   
     \item[Single-HMIM]  which works in the single-task setting.  
  \end{description}
  
  \item[Algorithmic:]  In the multiple task setting where the mechanism does not know the hierarchical information structure, we design an algorithm which, given agent reports as input,  outputs the hierarchical structure of the information as well as the most valuable information. Note that our algorithm is robust against agents with limited  (or no) information reporting biased (or totally random and useless) information.  This is used as part of the \emph{Learning-based Multi-HMIM} mechanism.
 
\end{description}

\subsection{Road Map}

The remainder of this section discusses related work; Section~\ref{sec:prelim} lays out the mathematical tools we use and can be referred back to when necessary.  Section~\ref{sec:framework} introduces both our model, the hierarchical information structure, and our mechanism design framework, the hierarchical mutual information paradigm (HMIP).  Section~\ref{sec:multi} and section~\ref{sec:single} apply the HMIP mechanism design framework into multi-task setting and single-task setting respectively. \gs{Section~\ref{sec:multi} and section~\ref{sec:single} can be read independently. --what is meant here?}

In the multi-task setting (Section~\ref{sec:multi}), we consider two situations: 1) the mechanism knows the information structure shared by agents but each agent is assigned a small number of tasks; 2) the mechanism does not know the information structure but each agent is assigned a large number of tasks such that the mechanism can learn the information structure. In the single-task setting (Section~\ref{sec:single}), we assume the mechanism know the information structure. 

Section~\ref{sec:discussion} concludes by discussing some potential applications and the robustness of our mechanisms.

\subsection{Related Work}

\paragraph{Model perspective} 
Prior work has modeled heterogeneous expertise where different agents receive a different number of signals \cite{goelrp09} or expertise is embedding in several dimensions~\cite{10.2307/2346806,zhou2012learning,welinder2010multidimensional,mccoy2017statistical}; however in these works lower expertise/effort along with a certain dimension only leads to a more noisy signal.  In contrast,  our model allows such signals to be systematically biased. 



\paragraph{Mechanism design perspective} The most related work with the current paper is \citet{prelec2017solution} which uses Bayesian Truth Serum~\cite{prelec2004bayesian} to incentivize agents to report their signal and selects the most surprising signal (measured by occurring more than its average prediction) as the final answer.  \citet{mccoy2017statistical} follow \citet{prelec2017solution} to propose a probabilistic model to learn the expertise of agents. \citet{riley2015mechanisms} compares the peer prediction decision rule (similar to \citet{prelec2017solution}) and the majority vote rule and exhibits cases where each outperforms the other. The current paper differs with \citet{prelec2017solution} in the model and assumptions as well as the possible applications. \citet{prelec2017solution} only focus on the single-task setting and assume that agents receive the signals endogenously (without effort). In contrast, this paper considers both single and multiple task settings and the model used in this paper handles both exogenous and indigenous signals.

The mechanism design framework in the current paper extends the information theoretic framework proposed in~\citet{2016arXiv160501021K}. \citet{DBLP:conf/sigecom/AgarwalMP017} propose a mechanism that works for the heterogeneous participants in the multi-task setting. \citet{mandal2016peer} consider the setting with heterogeneous tasks. They all do not assume the hierarchy of the information and cannot be applied to identify and elicit expertise.

\paragraph{Algorithmic perspective} Several works 
\cite{zhou2012learning,10.2307/2346806,welinder2010multidimensional,ghosh2011moderates} provide clever methods to learn the expertise as well as the ground truth of the crowdsourcing tasks. The algorithm in the current paper differs in two main aspects: (1) The current paper uses a different expertise model which can successfully capture the possibly hierarchical relationship between different information/expertise as well as the most valuable information; (2) the current paper combines the algorithm with an incentive mechanism that endogenously controls the quality and structure of the input, rather than making exogenous assumptions about the quality of the input.

\section{Mechanism Design Tools}\label{sec:prelim}

We use two key information theory ingredients in designing information elicitation mechanisms. The first ingredient is $f$-mutual information $MI^f(X;Y)$ which measures the amount of information crossing two random variables $X,Y$. For example, if $X$ is independent with $Y$---no information crosses $X$ and $Y$, $MI^f(X;Y)=0$. The second ingredient is proper scoring rule $PS(x,\mathbf{p})$ which measures the accuracy of the prediction $\mathbf{p}$ even we only have one sample $x$ of the outcome $X$. 

Both two ingredients have the information monotonicity property. If the information is measured by $f$-mutual information, any ``data processing'' on either of the random variables will decrease the amount of information crossing them. If the accuracy of a forecast is measured by a proper scoring rule, more information implies a more accurate forecast.

\subsection{$f$-mutual information}~\label{sec:fmutual}
$f$-divergence~\cite{ali1966general,csiszar2004information} $D_f:\Delta_{\Sigma}\times \Delta_{\Sigma}\rightarrow \mathbb{R}$ is a non-symmetric measure of the difference between distribution $\mathbf{p}\in \Delta_{\Sigma} $ and distribution $\mathbf{q}\in \Delta_{\Sigma} $ 
and is defined to be $D_f(\mathbf{p},\mathbf{q}):=\sum_{\sigma\in \Sigma}
\mathbf{p}(\sigma)f\left( \frac{\mathbf{q}(\sigma)}{\mathbf{p}(\sigma)}\right)$
where $f(\cdot)$ is a convex function and $f(1)=0$. Two commonly used $f$-divergences are KL divergence and total variation distance. Now we start to introduce $f$-mutual information.

Given two random variables $X,Y$, let $\mathbf{U}_{X,Y}$ and $\mathbf{V}_{X,Y}$ be two probability measures where $\mathbf{U}_{X,Y}$ is the joint distribution of $(X,Y)$ and $\mathbf{V}$ is the product of the marginal distributions of $X$ and $Y$. Formally, for every pair of $(x,y)$, $$\mathbf{U}_{X,Y}(X=x,Y=y)=\Pr[X=x,Y=y]\qquad \mathbf{V}_{X,Y}(X=x,Y=y)=\Pr[X=x]\Pr[Y=y].$$ 

If $\mathbf{U}_{X,Y}$ is very different with $\mathbf{V}_{X,Y}$, the mutual information between $X$ and $Y$ should be high since knowing $X$ changes the belief for $Y$ a lot. If $\mathbf{U}_{X,Y}$ equals to $\mathbf{V}_{X,Y}$, the mutual information between $X$ and $Y$ should be zero since $X$ is independent with $Y$. Intuitively, the ``distance'' between $\mathbf{U}_{X,Y}$ and $\mathbf{V}_{X,Y}$ represents the mutual information between them.

\begin{definition}[$f$-mutual information \cite{2016arXiv160501021K}]
The $f$-mutual information between $X$ and $Y$ is defined as $$MI^f(X;Y)=D_f(\mathbf{U}_{X,Y},\mathbf{V}_{X,Y})$$ where $D_f$ is $f$-divergence. 
\end{definition}

\begin{definition}[Conditional $f$-mutual information \cite{2016arXiv160501021K}]\label{def:cmi}
Given three random variables $X,Y,Z$, we define $MI^f(X;Y|Z)$ as $$\sum_z Pr[Z=z] MI^f(X;Y|Z=z)$$ where $MI^f(X;Y|Z=z):=MI^f(X';Y')$ where $Pr[X'=x,Y'=y]=Pr[X=x,Y=y|Z=z]$.
\end{definition}

Two examples of $f$-mutual information are Shannon mutual information~\cite{cover2006elements} (Choosing $f$-divergence as KL divergence) and $MI^{tvd}(X;Y):=\sum_{x,y}|\Pr[X=x,Y=y]-\Pr[X=x]\Pr[Y=y]|$ (Choosing $f$-divergence as Total Variation Distance).

\begin{lemma}[General data processing inequality / Information monotonicity \cite{2016arXiv160501021K}]\label{lem:imdpi} 
When $f$ is convex, $f$-mutual information $MI^f$ is symmetric $MI^f(X;Y)=MI^f(Y;X)$; non-negative and satisfies data processing equality: for any transition probability $M\in\mathbb{R}^{|\Sigma_X|\times |\Sigma_X|}$, when $Y$ is independent with $M(X)$ conditioning on $X$, $MI^f(M(X);Y)\leq MI^f(X;Y)$. 
\end{lemma}

\ifnum\fullversion=1

In addition to the above lemma, $f$-mutual information also satisfies the convexity defined following. 

\begin{lemma}[Convexity of $f$-mutual information \cite{2016arXiv160501021K}]\label{lemma:convexity}
Let $X_1,X_2,X,Y,Z,B_{\lambda}$ be random variables such that $X=\begin{cases}
X_1& B_{\lambda}=1\\
X_2& B_{\lambda}=0
\end{cases}$ where $B_{\lambda}$ is a Bernoulli variable that is independent of $X_0,X_1,Y,Z$, then 
$$MI^{f}(X;Y)\leq \lambda MI^{f}(X_1;Y)+(1-\lambda) MI^{f}(X_2;Y).$$
\end{lemma}

\else 

\fi




\subsection{Proper scoring rules}
Informally, a scoring rule measures the accuracy of the forecasts. Formally, a scoring rule~\cite{winkler1969scoring,gneiting2007strictly} $PS:  \Sigma \times \Delta_{\Sigma} \rightarrow \mathbb{R}$ takes in a signal $x \in \Sigma$  and a distribution over signals $\delta_{\Sigma} \in \Delta_{\Sigma}$ and outputs a real number. A scoring rule is \emph{proper} if, whenever the first input is drawn from a distribution $\delta_{\Sigma}$, then $\delta_{\Sigma}$ will maximize the expectation of $PS$ over all possible inputs in $\Delta_{\Sigma}$ to the second coordinate. A scoring rule is called \emph{strictly proper} if this maximum is unique. We will assume throughout that the scoring rules we use are strictly proper. Slightly abusing notation, we can extend a scoring rule to be $PS:  \Delta_{\Sigma} \times \Delta_{\Sigma} \rightarrow \mathbb{R}$  by simply taking $PS(\delta_{\Sigma}, \delta'_{\Sigma}) = \E_{x \leftarrow \delta_{\Sigma}}(x,  \delta'_{\Sigma})$.  \ifnum\fullversion=1 We note that this means that any proper scoring rule is linear in the first term.


\begin{example}[Log Scoring Rule~\cite{winkler1969scoring,gneiting2007strictly}]
Fix an outcome space $\Sigma$ for a signal $x$.  Let $\mathbf{q} \in \Delta_{\Sigma}$ be a reported distribution.
The Logarithmic Scoring Rule maps a signal and reported distribution to a payoff as follows:
$$L(x,\mathbf{q})=\log (\mathbf{q}(x)).$$

Let the signal $x$ be drawn from some random process with distribution $\mathbf{p} \in \Delta_\Sigma$.

Then the expected payoff of the Logarithmic Scoring Rule
$$ \E_{x \leftarrow \mathbf{p}}[L(x,\mathbf{q})]=\sum_{x}\mathbf{p}(x)\log \mathbf{q}(x)=L(\mathbf{p},\mathbf{q})$$

This value will be maximized if and only if $\mathbf{q}=\mathbf{p}$.

\end{example}

Intuitively, more information should imply a more accurate prediction. This intuition is valid when the accuracy is measured by a proper scoring rule. When predicting a random variable $Y$, assuming that all agents have a common prior, the agent who has more information will have higher prediction score when the prediction score is measured by a proper scoring rule. We denote the prediction of $Y$ conditioning on $X$ as $\Pr[\bm{Y}|X]:=(\Pr[Y=1|X],\Pr[Y=2|X],...,\Pr[Y=|\Sigma||X])\in \Delta_{\Sigma}$.



\begin{lemma}[Information monotonicity of proper scoring rules]\label{lem:psim}
Given any strictly proper scoring rule $PS$,  
$$\E_{X,Y,Z} PS(Y,\Pr[\bm{Y}|X,Z])\geq \E_{X,Y} PS(Y,\Pr[\bm{Y}|X]).$$
The equality holds if and only if $\Pr[\bm{Y}|X=x,Z=z]=\Pr[\bm{Y}|X=x]$ for all $(x,z)$ where $\Pr[X=x,Z=z]>0$.
\end{lemma}

We defer the proof to the appendix. 


\fi

\section{Model and Mechanism Design Framework}\label{sec:framework}



\begin{figure}[h]
\centering
\includegraphics[width=0.8\linewidth]{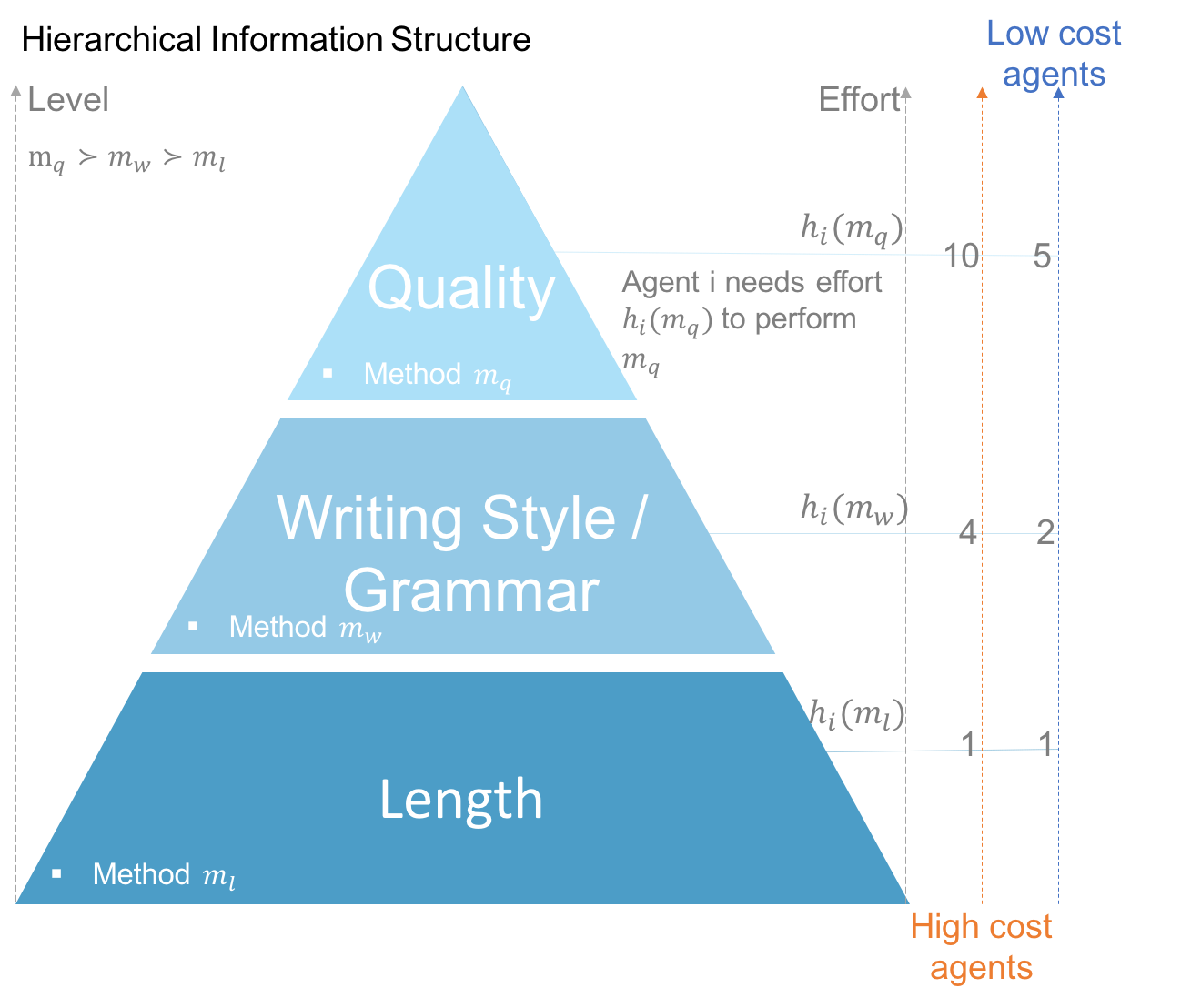}
\caption{An illustration of the hierarchical information structure in the peer grading process.
 }
\label{fig:peer}
\end{figure}

In this section, we will define the hierarchical information structure and provide a mechanism design framework that helps design mechanisms which elicit the hierarchical information.  Section~\ref{sec:model} defines the information model; Section~\ref{sec:mechanism-framework} defines our mechanism framework; and Section~\ref{sec:mechanism-analysis} analyzes the framework.  We will use the peer grading process (Figure~\ref{fig:peer}) as a running example to throughout this section. 









\subsection{Hierarchical  Information Structure} \label{sec:model}


There are $n$ agents and one task. The agents have a finite set $M$ of methods to perform on the task based on the task's attributes $\mathbf{a}\in A$ where $a$ is a random (possibly high dimensional) vector drawn from a distribution $Q_A\in \Delta_{A}$. Each method $m:A\mapsto \Sigma_m$ maps the attributes $\mathbf{a}\in A$ to a signal $m(\mathbf{a})$ from a finite set $\Sigma_m$. We now introduce our peer grading example. 

\begin{note}
10 evaluators are asked to judge one essay. The essay has eight possible attributes: $\mathbf{a}=(q_i,w_j,l_k), i,j,k\in \{0,1\}$. $(q_1,w_0,l_1)$ means the essay has (good quality, bad writing, long length); $(q_0,w_1,l_0)$ means the essay has (bad quality, good writing, short length). The distribution over the attributes space $Q_A$ is defined as:
\gs{note that I added the star's following.  Please make consistent.}
\begin{center}
     {
\begin{tabular}{c|c|c|c}
  $Q_A((q_0,w_0, *))$   & $Q_A((q_0,w_1, *))$ & $Q_A((q_1,w_0, *))$ & $Q_A((q_1,w_1, *))$ \\
  \hline
  0.4   & 0.1 & 0.1 & 0.4 \\
\end{tabular}

}%
     \end{center}
With this distribution, an essay of good quality usually has good writing as well. Moreover,
we assume the essay's length is independent with the essay's quality and writing and an essay has long length with probability 0.5.  That is: $Q_A((q_i,w_j,l_1))=Q_A((q_i,w_j,*))*0.5, Q_A((q_i,w_j,l_0))=Q_A((q_i,w_j,*))*0.5$.

Each evaluator can perform three methods:  $m_l(\mathbf{a})$,  $m_w(\mathbf{a})$, and  $m_q(\mathbf{a})$ which are, respectively, (possibly noisy) signals about the essay's length; writing style and grammar; and  quality. $\Sigma_l=\Sigma_w=\Sigma_q=\{\smiley,\frownie\}$.
\end{note}

\yk{I wrote a new version}

We define $\psi_{i}^m(\mathbf{a})$ as agent $i$'s received output by performing $m$ on attributes $\mathbf{a}$. Different agents may receive different signals by performing the same method on the same attributes. But we assume the distribution is symmetric/homogeneous in the sense that for any permutation $\pi:[n]\mapsto [n]$, the probability that $\psi_{1}^m(\mathbf{a})=\sigma_1$, $\psi_{2}^m(\mathbf{a})=\sigma_2$, ... $\psi_{n}^m(\mathbf{a})=\sigma_n$ equals the probability that $\psi_{\pi(1)}^m(\mathbf{a})=\sigma_1$, $\psi_{\pi(2)}^m(\mathbf{a})=\sigma_2$, ... $\psi_{\pi(n)}^m(\mathbf{a})=\sigma_n$. We also assume that each agent performs methods independently (see (\ref{eq:ind})). When the attributes $\mathbf{a}$ is drawn from a distribution $Q_A$, we can define define $\Psi_i^m$ as agent $i$'s received output by performing $m$ on a random attributes $\mathbf{a}$ that is drawn from a distribution $Q_A$. Analogously, we define a random variable $\Psi_{-i}^m$ as an arbitrary agent $j\neq$'s received output by performing $m$ on a random attributes $\mathbf{a}$ that is drawn from a distribution $Q_A$. This definition is well-defined since we have assumed the distribution is symmetric. We define prior $Q$ as a joint distribution over all $\{\Psi_i^m\}_{i\in[n],m\in M}$.

\begin{note}
Conditioning on the attributes of the essay $\mathbf{a}=(q_i,w_j,l_k), i,j,k\in \{0,1\}$, for each method $m$, each agent will receive $\psi_{i}^m(\mathbf{a})=\smiley$ with probability $p_{m,\mathbf{a}}$ independently by performing $m$. That is, agents' received signals by performing $m$ is a Binomial distribution $B(n=10,p_{m,\mathbf{a}})$. \footnote{To give a concrete example, we use the Binomial distribution here. In fact, we only need the distribution to be symmetric.}\\


\begin{tabular}{c|c|c}
     & good quality essay\footnote{This means $\mathbf{a}=(q_1,*,*)$} & bad quality essay\\
     \hline
     $\Pr[m_q(\mathbf{a})=\smiley]$ & 70\%  & 30\%
\end{tabular}\\

This means conditioning on the essay having good quality, the distribution over agents' received quality signals by performing $m_q$ is $Q_{m_q,(q_1,*,*)}=B(10,0.7)$; while conditioning on the essay having good quality, the distribution is $Q_{m_q,(q_0,*,*)}=B(10,0.3)$. Similarly, we have\\

\begin{tabular}{c|c|c}
     & good writing essay & bad writing essay\\
     \hline
     $\Pr[m_w(\mathbf{a})=\smiley]$ & 90\%  & 10\%
\end{tabular}\\

\begin{tabular}{c|c|c}
     & long essay & short essay\\
     \hline
     $\Pr[m_l(\mathbf{a})=\smiley]$ & 100\%  & 0\%
\end{tabular}\\

Note that the cheap length signal is noiseless. We also assume that fixing the attributes, every agent performs the different methods independently. That is, when $\mathbf{a}=(q_1,w_1,l_1)$ 
\begin{align}\label{eq:ind}
    \Pr\left(\Psi_i^{m_l}(\mathbf{a})=\smiley,\Psi_i^{m_w}(\mathbf{a})=\smiley,\Psi_i^{m_q}(\mathbf{a})=\smiley\right)=0.7*0.9*1.
\end{align}

With the above set up, the probability that agent $i$ receives a $\smiley$ writing signal and agent $j$ receives a $\smiley$ quality signal will be

    $\Pr[\Psi_i^{m_w}=\smiley,\Psi_j^{m_q}=\smiley]=0.4*0.1*0.3+0.1*0.3*0.9+0.1*0.7*0.1 + 0.4*0.7*0.9=0.298$.

\end{note} We define a partial order on the methods. We say ${m_1}\succeq {m_2}$---the level of $m_1$ is higher than that of $m_2$---if method $m_1$ cannot be performed without performing $m_2$.  By ${m_1}\succ {m_2}$ we mean ${m_1}\succeq {m_2}$ but ${m_2}\not\succeq {m_1}$. Note that the partial order $\succ$ is transitive---$m_1\succ m_2, m_2\succ m_3\Rightarrow m_1\succ m_3$. Each agent $i$ needs effort $h_i(m)>0$ to perform method $m$ and when she spends effort $h_i(m)$ to perform $m$, the methods that have lower levels than $m$ are performed as well without additional effort. We assume, as is natural, that $h_i(m)$ is an increasing function, that is, $h_i(m_1)\geq  h_i(m_2)$ when $m_1\succeq m_2$. The higher the level of the method an agent performs, the more effort she must invest.  However, it may be the case that some agents (low cost agents) can perform methods more economically than others (high cost agents). The partial order definition is essentially our key assumption (Assumption~\ref{assume:hier}). 

\begin{note}
$m_q\succ m_w\succ m_a$. Among the 10 evaluators, there are 2 low cost evaluators who need $1,2,5$ effort to perform $m_l,m_w,m_q$ respectively. There are 8 high cost evaluators who need  $1,4,10$ effort to perform $m_l,m_w,m_q$ respectively (Figure~\ref{fig:peer}). Based on the partial order definition, when an evaluator spends sufficient effort to perform $m_q$ and obtains the quality signal, she also obtains the length and writing signals without additional effort, which is natural in real life.
\end{note}
We assume agents share a hierarchical information structure and allow agents to have different priors $Q$\footnote{To ease the presentation of the example, in our peer grading example, we assume agents share the same prior $Q$.}. We will design mechanisms that incentivize agents to invest efforts based on their costs and report honestly.


\subsection{Mechanism Design Framework}
\label{sec:mechanism-framework}

We start by introducing the formal definition of a mechanism. 

\begin{definition}[Mechanism]
We define a \emph{mechanism} $\mathcal{M}$ for $n$ agents as a tuple $\mathcal{M}:=(R,S)$ where $R$ is a set of all possible reports the mechanism allows, and $S:R^n\mapsto \mathbb{R}^n$ is a mapping from all agents' reports to each agent's payment.
\end{definition}

We will extend the Mutual Information Paradigm~\cite{2016arXiv160501021K} to our Hierarchical Mutual Information Paradigm that handles the hierarchical information structure. 

\paragraph{Mutual Information Paradigm~\cite{2016arXiv160501021K}}\citet{2016arXiv160501021K} provide a Mutual Information Paradigm for Peer Prediction Mechanisms: each agent $i$ is paid the mutual information between her information and her peers' information---

$$MI(\text{her information};\text{her peers' information}).$$

By picking $f$-mutual information $MI^f$ that satisfies data processing equality (Section~\ref{sec:fmutual}), no agent can obtain strict benefit by lying since intuitively the amount of information each agent has will not increase no matter what kind of strategy she applies to her information.

We can naturally extend the Mutual Information Paradigm for Peer Prediction Mechanisms to the hierarchical model by paying agent $i$ $$ MI^f(\text{her information};\{\Psi^m_{-i}\}_{m\in M})\footnote{Recall that $(\Psi_1^m,\Psi_2^m,...,\Psi_n^m)$ are the random signals agents receive by performing method $m$ on the same random attributes.} .$$


This idea has a severe drawback: sometimes low level information has very large correlation with the high level information. In this case, $ MI^f(\text{her information};\{\Psi^m_{-i}\}_{m\in M})$ will pay low level information nearly as much as  high level information; and so agents will lack incentive to perform high level methods.



To solve the above problem, we pay agents method by method. For each $m$, we only value the ``information gain'' in the sense that we pay each agent $i$ the mutual information between her information and the method $m$'s information \emph{conditioning on} the information output by the methods are lower than $m$. 

Formally, we chose a payment scale $\alpha_m$ for each $m$ and pay each agent $i$ $$\sum_{m}\alpha_m MI^f(\text{her information};\Psi^m_{-i}|\{\Psi^{m'}_{-i}\}_{m'\prec m}).$$

In our actual paradigm, we hope to pay each agent $i$ using the above payment when the mechanism has access to all levels of honest information provided by other agents.

\begin{note}
In the peer grading example, the information about the writing style / grammar may already have a very high correlation with the quality of the essay. With the above concrete set up, we are ready to calculate the (conditional) Shannon mutual information (Euler number base) between agent $i$'s received signals and agent $j$'s received signals. For example, the $2\times 2$ entry is the mutual information between agent $i$'s received length signal, writing signal by performing method $m_w$ and agent $j$'s writing signal, conditioning on agent $j$'s length signal, which is 0.2259. We calculate the values by first calculating the joint distribution over 6 random variables---agent $i$'s length, writing, quality signals and agent $j$'s length, writing, quality signals.

We show the values in the following table and defer the calculation to \ifnum\fullversion=1 Appendix~\ref{sec:cal}\else the full version\fi. According to the information monotonicity, for each column, the values increase from bottom to top. 

\begin{center}
    
\gs{I think the second column should shift to the last column and we should say: The last column is just the sum of the previous three."}

{\footnotesize
\begin{tabular}{ c | c| c | c|  c  } 
    \hline
     \diagbox[innerwidth=3.5cm]{agent $i$'s}{$MI(\cdot;\cdot)$}{agent $j$'s}  & length & writing$|$length \footnote{x$|$y means x conditioning on y.} & quality $|$ writing, length &  length, writing, quality\footnote{Since we use Shannon mutual information which satisfies chain rule, the last column is the sum of the previous columns.}\\
     \hline
    length, writing, quality   & 0.6931 & 0.2259 & \textbf{0.0115}& \textbf{0.9305}\\
    \hline
     length, writing  & 0.6931 & 0.2218 & \textbf{0.0041}  & \textbf{0.9190}\\
    \hline
     length & 0.6931 & 0 & 0 & 0.6931 \\ 
    \hline
\end{tabular}
}
\end{center}
\vspace{3pt}
Even though performing the quality method provides the information that has the highest mutual information 0.9305 with other agents' information, performing writing method already outputs information that has $0.9190\approx 0.9305*0.98$ mutual information with other agents' information. 

In this case, what we really value is the additional quality of information after conditioning on the information of cheap signals like writing style / grammar. In other words, we value the information about an essay which has a high quality but is written carelessly (or low quality but impeccable prose). 

Each agent, performing the writing method only has 0.0041 mutual information with other agents' quality signal conditioning on other agents' writing and length signals while performing the quality method has $0.0115\approx 0.0041*2.80$ conditional mutual information. 

Looking ahead, we seek to pay each evaluator $i$ by:  
\begin{align*}\label{eq11} \numberthis
    &\alpha_l MI^f(\text{her information};\text{agent $j$'s length signal})\\
    +&\alpha_w MI^f(\text{her information};\text{agent $j$'s writing signal}|\text{agent $j$'s length signal})\\
    +&\alpha_q MI^f(\text{her information};\text{agent $j$'s quality signal}|\text{agent $j$'s length and writing signal}).\end{align*}
where $\alpha_q$ is set to be rather larger than $\alpha_l$ and $\alpha_w$.

\end{note}




\paragraph{Hierarchical Mutual Information Paradigm (HMIP($MI^f$,$\{\alpha_m\}_m$))} We now present our hierarchical Mutual Information Paradigm.  We emphasize that this is not a mechanism that can be run. Instead we engage in the wishful thinking that the reports of the agents are distributions rather than draws from the distribution.  Of course, this will never happen.  Nonetheless, we will show that using the HMIP paradigm we can design actual mechanisms in both the multiple-task setting (Section~\ref{sec:multi}) and the single-task setting (Section~\ref{sec:single}).

The paradigm requires as parameters a payment scale $\alpha_m \in \mathbb{R}_{\geq 0}$ for each method $m$.

\begin{description}

\item[Report] For each agent $i$, for each $m \in M$, she is asked to optionally provide the random signal $\Psi_i^m$. We denote the set of methods whose outputs are reported by agent $i$ as $\methodset_i$ and the actual random signal she reports for each $\ell\in\methodset_i$ as $\hat{\Psi}_i^{\ell}$. 
\item[Payment/Information Score] We define $\methodset_{-i}$ as $\bigcup_{j\neq i} M_j$. For each $m\in \methodset_{-i}$, we arbitrarily pick an agent $j\neq i$ who provides method $m$'s output and denote his report for method $m$'s output as $\hat{\Psi}_{-i}^m$. 

Agent $i$ is paid by her information score
$$ \sum_{m\in\methodset_{-i}}\alpha_m MI^{f}(\{\hat{\Psi}_i^{\ell}\}_{\ell\in\methodset_i};\hat{\Psi}_{-i}^m|\{\hat{\Psi}_{-i}^{m'}\}_{m'\prec m,m'\in \methodset_{-i}})$$ 
\end{description}

\paragraph{Resolving ``wishful thinking''} HMIP pays agents according to the information measure. The calculation of the information measure requires the knowledge of the prior, i.e., the joint distribution which is unrealistic in practice. To resolve this ``wishful thinking'', a key observation is that paying agents an unbiased estimator of the information measure is sufficient when we assume agents are expected utility maximizers. To construct an unbiased estimator of the information measure using agents' reports, different settings have different techniques. In the multi-task setting, either we ask a large number of questions to estimate the prior and use the prior to calculate the information measure (Learning-based multi-HMIM), or we ask a small number questions but require the knowledge of the structure and use a special $f$-mutual information, $MI^{tvd}$ (Multi-HMIM). In the single-task setting (Single-HMIM), we ask agents their posteriors (e.g. what percentage of your peers say yes?) and construct the estimator using both the first order information (e.g.,\ Y/N) and the second order information (e.g.,\ 80\% Yes). Thus, although all proposed mechanisms are based on the HMIP, they are all detail free in the sense that they do not need any priori knowledge of the distributions (nor wishful thinking). 

\subsection{Analysis of HMIP}  \label{sec:mechanism-analysis}
For each agent $i$, we define her utility as her payment minus her effort. 

\begin{definition}[Strategy]
We define the \emph{effort strategy} of each agent $i$ as a mapping $e_i$ from her priors to a probability distribution over the methods she will perform. We define the \emph{report strategy} of each agent $i$ as a mapping $s_i$ from her received information to a probability distribution over $R$.
\end{definition}

\gs{Seems like the following definition should be moved first, and then prudent strategy should be defined}
\begin{definition}[Amount of information in HMIP]\label{def:aoihmip}
In HMIP, for agent $i$, the amount of information acquired with method $m_i$ is defined as 
$$AOI(m_i,\text{HMIP}(MI^f,\{\alpha_m\}_m)):=\sum_{m\in M}\alpha_m MI^{f}(\{\Psi_i^{\ell}\}_{\ell\preceq{m_i}};\Psi_{-i}^m|\{\Psi_{-i}^{m'}\}_{m'\prec m}).$$
\end{definition}

\gs{say what this is in words}

\gs{Why do we have this?: 
We say the definition of amount of information is valid if it is an increasing function of the level of the method. The information monotonicity of $MI^f$ implies that the amount of information defined here is valid.} 

We have already give the example of the amount of information in (\ref{eq11}). Later in the proof of Theorem~\ref{thm:HMIP}, we will see the amount of information is also the optimal payment of agent $i$ who performs method $m_i$ when HMIP has access to all levels of honest signals reported by other agents. 

An especially desirable strategy in HMIP is a \emph{prudent strategy}. Informally, agents play a prudent strategy if they (a) choose the method they perform to maximize their utility---trading off the amount of information acquired with the effort it costs; (b) report all received information honestly.





\begin{definition}[Prudent strategy in HMIP]\label{def:prudentstrategy}
For each agent $i$, we say she plays a prudent strategy in HMIP($MI^f$,$\{\alpha_m\}_m$) if she chooses to (a) perform method $m_i^*$ such that 
$$m_i^*\in \argmax_{m_i}\left(AOI(m_i,\text{HMIP}(MI^f,\{\alpha_m\}_m)) -h_i(m_i)\right); and$$ (b) reports all received information honestly. 
\end{definition}


\begin{definition}[Truthful strategy in HMIP]\label{def:truthfulstrategy}
We say an agent plays truthful strategy if she always reports her received information honestly. 
\end{definition}

A truthful strategy is a special report strategy. An agent can play any effort strategy and truthful strategy simultaneously. If an agent invests no effort and reports nothing or meaningless information, she is still considered as playing truthful strategy.  

\paragraph{Mechanism design goals} A mechanism $\mathcal{M}$ is (strictly)  \emph{\faithful} if for each agent, when she believes everyone else plays their prudent strategy, she can (strictly) maximize her expected utility by playing a prudent strategy as well. A mechanism $\mathcal{M}$ is \emph{dominant truthful} if for each agent, regardless of other agents' strategies, she can maximize her expected utility by playing a pure effort strategy and truthful strategy. 

\yk{we can change the word for faithful by change it in the main file. }


The dominant truthful property is incomparable with the \faithful property. A flat payment scheme is dominant truthful but not \faithful since investing no effort and reporting nothing is also considered as a pure effort and truthful strategy. The \faithful property is desirable since it encourages low cost agents to invest high level effort and high cost agents to invest low level effort, and incentivizes them to report honestly as well. 

In order to design \faithful mechanism, the coefficients $\{\alpha_m\}_m$ should be chosen appropriately. \gs{removed:  set prudently.} If the coefficients $\{\alpha_m\}_m$ are too low, no agent will be incentivized to invest the highest level method. In this case, the mechanism cannot access all levels of information to encourage agents to play prudent strategy. Thus, we will assume the coefficients are chosen  such that it is worthwhile for at least two agents to invest the highest level method. 

We say a method $m$ is \emph{maximal} if there does not exist $m'\neq m\in  M$ such that $m'\succ m$.  

\gs{Could say that it is "$m$-supporting" if BLAH -- for a particular $m$?}

\begin{definition}[\faithful coefficients for HMIP]\label{def:prudentcoefficients}
Given the priors $\{Q_m\}_m$, we say the coefficients $\{\alpha_m\}_m$ are \faithful for HMIP($MI^f$,$\{\alpha_m\}_m$)  if given the coefficients $\{\alpha_m\}_m$, for every maximal $m$, there exists at least \textbf{two} agents whose prudent strategy in HMIP($MI^f$,$\{\alpha_m\}_m$)  is performing method $m$. 
\end{definition}

This is a weak requirement since we only need to set sufficiently high coefficients to incentivize \textbf{two} low cost agents such that for each agent (including one of the low cost agent), she will believe there exists a low cost agent who will be incentivized to report all levels of information. \faithful coefficients exist since we can always set the coefficient of the highest level information sufficiently high and the coefficients of other levels arbitrarily close to zero such that agents will be incentivized to invest the highest level effort. We use our peer grading example to show how to set \faithful coefficients. With our example, we will see we can always set the optimal \faithful coefficients that minimize the mechanism's cost by solving a linear programming.

\begin{note}
In our example, the 2 low cost agents need efforts 1,2,5 to perform $m_l,m_w,m_q$ respectively and 8 high cost agents need efforts 1,4,10. With the above set up, we need
$\alpha_q*0.0115+\alpha_w*0.2259+\alpha_l*0.6931-5 > \max\{\alpha_q*0.0041+\alpha_w*0.2218+\alpha_l*0.6931-2,\alpha_l*0.6931-1,0\}$ to make the coefficients \faithful and we also want to minimize the mechanism's cost which is 
{\small

\begin{align*}
    &2*(\alpha_q*0.0115+\alpha_w*0.2259+\alpha_l*0.6931)\\
    &+8*\begin{cases}
v_q:=\alpha_q*0.0115+\alpha_w*0.2259+\alpha_l*0.6931& \text{if $v_q-10\geq v_w-4,v_l-1,0$}\\
v_w:=\alpha_q*0.0041+\alpha_w*0.2218+\alpha_l*0.6931& \text{if $v_w-4\geq v_q-10,v_l-1,0$}\\
v_l:=\alpha_l*0.6931& \text{if $v_l-1\geq v_w-4,v_q-10,0$}\\
0& \text{otherwise}\\
\end{cases}
\end{align*}

}

After solving this linear programming, the optimal solution is around $\alpha_l,\alpha_w,\alpha_q =\epsilon \footnote{$\epsilon$ is an arbitrarily small positive real number, we need $\epsilon$ since we want agents will be incentivized to report the length signal as well.},0.5562,423.8571$  and the amount of information for performing $m_l$,$m_w$,and $m_q$ are $O(\epsilon)$, $1.86+O(\epsilon)$, and $5+O(\epsilon)$ respectively. The minimal cost is $2*(5+O(\epsilon))=10+O(\epsilon)$. 

\gs{we should move this to the appendix.  Instead summarize it here.  The more important thing is to say that without eliciting multiple levels, NO payments support any agents revealing quality in ANY equilibrium.}

\gs{Also, this is not clear what is meant here}


\end{note}

\gs{\faithful-$m$-supporting???}

\begin{theorem}\label{thm:HMIP}

Given a convex function $f$, HMIP($MI^f$,$\{\alpha_m\}_m$) is dominant truthful; moreover, when $\{\alpha_m\}_m$ are \faithful for HMIP($MI^f$,$\{\alpha_m\}_m$),  HMIP($MI^f$,$\{\alpha_m\}_m$) is \faithful and dominant truthful. 

\end{theorem}

The proof of the theorem uses the information monotonicity of $MI^{f}$. The key observation in the proof is that \emph{applying any strategy to the information is essentially data processing and thus erodes information}.




\begin{proof}[Proof for Theorem~\ref{thm:HMIP}]



In order to show the dominant truthful property, we will show for each agent, fixing any other agents' strategies, she can maximize her payment as well as her utility by reporting her received information honestly. The information monotonicity property of $f$-mutual information $MI^{f}$ (Fact~\ref{lem:imdpi}) says any data processing decreases the (conditional) mutual information. For each $m\in \methodset_{-i}$, fixing the strategies other agents use, the distribution of $\hat{\Psi}_{-i}^m$, whose randomness comes from random variable $\Psi_{-i}^m$ and the agents' strategies, is also fixed. Any strategy (data processing) agent $i$ applies to her received signals decreases $$MI^{f}(\text{her received signals};\hat{\Psi}_{-i}^m|\{\hat{\Psi}_{-i}^{m'}\}_{m'\prec m,m'\in \methodset_{-i}}).$$ Thus, for agent $i$, honestly reporting her received signals maximizes her payment no matter what strategies other agents use.


We start to show HMIP is \faithful when the coefficients are \faithful. When the coefficients are \faithful, for every agent $i$, when she believes everyone else plays a prudent strategy, she will believe for each $m$, there exists an agent $j(m)\neq i$ who reports $\{\Psi_j^{\ell}\}_{\ell\preceq{m}}$ \gs{there is a mis-match of variables here}  to the mechanism. Thus, $\methodset_{-i}=M$. In this case, agent $i$'s optimal payment for performing method $m_i$ will be $$\sum_{m\in M}\alpha_m MI^{f}(\{\Psi_i^{\ell}\}_{\ell\preceq{m_i}};\Psi_{j}^m|\{\Psi_{j}^{m'}\}_{m'\prec m})=\sum_{m\in M}\alpha_m MI^{f}(\{\Psi_i^{\ell}\}_{\ell\preceq{m_i}};\Psi_{-i}^m|\{\Psi_{-i}^{m'}\}_{m'\prec m})$$ which is the amount of information acquired with method $m_i$ (Definition~\ref{def:aoihmip}) and can be obtained by honestly reporting all received information. The optimality of the payment is due to the information monotonicity of $MI^f$. In this case, her optimal strategy is her prudent strategy. Therefore, HMIP is \faithful.

\end{proof}



HMIP provides a framework to design information elicitation mechanisms for our hierarchical information model. To apply the HMIP framework in different settings, it remains to design the report requirement for agents and to use agents' reports to calculate the (conditional) mutual information without underlying distributions. We apply HMIP in both the multi-task setting (Section~\ref{sec:multi}) and the single-task setting (Section~\ref{sec:single}).


\section{Multi-task Setting}\label{sec:multi}
In this section, we will apply the HMIP framework in the multi-task setting where each agent receives a random batch of \emph{a priori similar} tasks\ifnum\fullversion=0: tasks are i.i.d samples for all agents. For every agent, before she invests any effort, for each $m$, for all tasks, she has the same prior belief for the signals she and other agents will receive by performing $m$.\else.\fi

\ifnum\fullversion=1

\subsection{Backgrounds and Assumptions}
In multi-task setting, the major challenge solved in previous peer prediction literature is that agents may ``get something for nothing'' by always answering the same answer (e.g. always saying good in peer grading).

In the setting where agents are assigned $\geq 2$ tasks, \citet{dasgupta2013crowdsourced,2016arXiv160501021K,2016arXiv160303151S} solve this challenge by assuming agents are homogeneous and rewarding agents not only for their agreements but also for the diversity of their answers. If an agent answers the same answer all the time (no diversity), she will be paid nothing. \citet{2016arXiv160501021K} show that this idea essentially means rewarding each agent $MI^{tvd}(\text{her information};\text{her peer's information}).$


When agents are heterogeneous, \citet{mandal2016peer} ask agents to answer a sufficient number of tasks and then classify their answers into different clusters to learn their levels and pay them.




\begin{assumption}[a priori similar]\label{assume:a priori}
All tasks are a priori similar for all agents. That is, tasks are i.i.d samples for all agents. For every agent, before she invests any effort, for each $m$, for all tasks, she has the same prior belief for the signals she and other agents will receive by performing $m$.
\end{assumption}

\else

\fi


Prior work~\cite{2016arXiv160303151S,dasgupta2013crowdsourced,2016arXiv160501021K} also makes this assumption; however, in their setting, it is much stronger than in ours.   For example, it insists that the only ``signal'' included in a prompt is for the correct answer.  In reality, some false answers are more appealing than others (see Example~\ref{eg:state} where Kansas is an unlikely answer).  In our model, these appealing false answers can be modeled as ``cheap'' information instead of being assumed away. 

Note that in the multi-task setting, we allow agents to have different priors and only require that for every agent, her prior satisfies our assumptions.

\subsection{Known Information Structure and a Small Number of tasks}\label{subsec:small}

In order to avoid agents ``getting something for nothing'' by reporting the cheap signals instead of the expensive signals (e.g. giving a high quality grade when there are no typos in Example~\ref{eg:peer}), we reward agents the information score of expensive signals according to not only their agreements but also the diversity of their answers \emph{conditioning on the tasks which have the same cheap signals. (e.g. the essays which all have no typos)}. We will show this idea is essentially the application of HMIP framework when $MI^f$ is chosen to be $MI^{tvd}$. 

\ifnum\fullversion=1

\begin{assumption}[Positively correlated signals]\label{assume:pc}
We assume that for every method $m$, each agent $i$, every $\sigma\neq \sigma'$, every possible $\{\sigma^{m'}\}_{m'\prec m}$, every subset $M'\subset \{m'|m'\prec m\}$,  $\Psi^m_{-i}$ is \emph{positively correlated} with $\Psi_i^m=\sigma$: $$\Pr[\Psi^m_{-i}=\sigma|\Psi_i^m=\sigma]>\Pr[\Psi^m_{-i}=\sigma],$$ $$\Pr[\Psi^m_{-i}=\sigma|\Psi_i^m=\sigma']<\Pr[\Psi^m_{-i}=\sigma],$$ conditioning on $\{\Psi_{-i}^{m'}\}_{m'\in M'}=\{\sigma^{m'}\}_{m'\in M}$.
\end{assumption}

\citet{dasgupta2013crowdsourced} and \citet{2016arXiv160303151S} both make this assumption as well. It means that receiving $\sigma$ by performing $m$ will increase each agent's belief for how many other agents receive $\sigma$ by performing $m$. It is a substantially weaker assumption than that agents always believe they are in the majority. 
\begin{note}
In the peer grading example, this assumption means that for every agent, receiving $\smiley$ for the quality signal will increase her belief for the probability other agents receive $\smiley$ for the quality signal. 
\end{note}

\begin{assumption}[Conditional independence]\label{assume:id}
For each agent $i$ who performs method $m_i$, we assume that for every possible $\{\sigma^{m'}\}_{m'\prec m}$, every subset $M'\subset \{m'|m'\prec m\}$, for each $m\preceq  m_i$, $\Psi_i^m$ contains all information agent $i$ has that is related to $\Psi^m_{-i}$, in other words, conditioning on $\Psi_i^m$, $\{\Psi_i^{m'}\}_{m'\preceq  m_i,m'\neq m}$ are independent with $\Psi_{-i}^m$, conditioning on $\{\Psi_{-i}^{m'}\}_{m'\in M'}=\{\sigma^{m'}\}_{m'\in M'}$\footnote{Note that if agents receive the same signal by performing the same method, both Assumption~\ref{assume:pc} and Assumption~\ref{assume:id} will hold.}.
\end{assumption}
\begin{note}
In the peer grading example, this assumption means that for every agent, if she has already thought the writing is good, her quality signal will not affect her opinion for the writing. 
\end{note}
With this assumption, when an agent needs to report her information that is related to $\Psi^m_{-i}$, assuming she has already performed method $m$, it's sufficient for her to only report $\Psi_i^m$.

\else

We assume agents' signals are \emph{positively correlated}. In the peer grading example, this assumption means that for every agent, receiving $\smiley$ for the quality signal will increase her belief for the probability other agents receive $\smiley$ for the quality signal. We defer the formal definition to the full version. 

\fi

\paragraph{Multi-task Hierarchical Mutual Information Mechanism (Multi-HMIM($\{\alpha_m\}_m$))}\begin{description}
\item [Report] Each agent $i$ is assigned a random batch of tasks (at least two). For each task $t$ which is assigned to agent $i$, she is asked to report both the method $m_i(t)$ she performed on task $t$ and method $m_i(t)$'s output ${\psi}_i^{{m}_i(t)}(t)$; for each $m \neq m_i(t)$, agent $i$ is asked to optionally report her signal ${\psi}_i^m(t)$. We denote her actual report for her performed method and signal for every method $m$ by $\hat{m}_i(t)$ and $\hat{\psi}_i^m(t)$ respectively.


\item [Information Score] For each method $m$, the mechanism collects agent $i$'s method $m$ signals and records them via a $T$ dimensional vector $\bm{\hat{\psi}}_i^m$.    

The $t^{th}$ coordinate of $\bm{\hat{\psi}}_i^m$ is $\begin{cases}
\hat{\psi}_i^m(t), & \text{if agent $i$ provides the }\\
 & \text{method $m$'s output $\hat{\psi}_i^m(t)$}\\
  & \text{for task $t$;}\\
 & \\
\emptyset,& \text{otherwise}
\end{cases}$

We define $\bm{\hat{\psi}}_{-i}^m$ as a vector where the $t^{th}$ coordinate of $\bm{\hat{\psi}}_{-i}^m$ is $$\begin{cases}
\hat{\psi}_{-i}^m(t), & \text{we arbitrarily pick an agent (who is not agent $i$)  }\\
  &\text{whose performed method is $\succeq m$ for task $t$ } \\
  &\text{and provides method's $m$'s output for task $t$;}\\
  &\text{we denote his report by $\hat{\psi}_{-i}^m(t)$; } \\
  & \\
\emptyset,&\text{such agent does not exist}
\end{cases}$$

Agent $i$ is paid by her information score $$\sum_{m}2\alpha_m Corr(\bm{\hat{\psi}}_i^m;\bm{\hat{\psi}}_{-i}^m|\bm{\{\hat{\psi}}_{-i}^{m'}\}_{m'\prec m})$$ and $Corr(\cdot)$\footnote{$Corr(\cdot;\cdot)$ is essentially the same concept as the payment schemes in \citet{dasgupta2013crowdsourced,2016arXiv160501021K,2016arXiv160303151S}. $Corr(\cdot;\cdot|\cdot)$ is a new concept in this paper.} is a random algorithm defined in Algorithm~\ref{corr}. 

\end{description}

We design $Corr(\bm{\hat{\psi}}_i^m;\bm{\hat{\psi}}_{-i}^m|\bm{\{\hat{\psi}}_{-i}^{m'}\}_{m'\prec m})$ to be an unbiased estimator of $MI^{tvd} ({\hat{\Psi}}_i^m;{\hat{\Psi}}_{-i}^m|{\{\hat{\Psi}}_{-i}^{m'}\}_{m'\prec m})$\footnote{In the current paper, $\bm{\hat{\psi}}_i^m$ means vector, ${\hat{\Psi}}_i^m$ means random variable.} if ${\hat{\Psi}}_i^m$ and ${\hat{\Psi}}_{-i}^m$ are \emph{positively correlated}. Thus, in Multi-HMIM, agents are essentially paid based on the (conditional) mutual information by picking a special $f$-mutual information---$MI^{tvd}$, if agents are honest since we have assumed that agents' honest signals are positively correlated. This makes our Multi-HMIM a special case of HMIP framework.

\ifnum\fullversion =1

\begin{definition}[Amount of information in Multi-HMIM]
In Multi-HMIM, when agent $i$ performs method $m_i$, the amount of information acquired with the effort is defined as 

$$AOI(m_i,\text{Multi-HMIM}(\{\alpha_m\}_m)):=\sum_{t\in[T]}\max_{f_m:\Pi_{\ell\preceq m_i}\Sigma_{\ell}\mapsto\Sigma_m}\alpha_m MI^{tvd}(f_m(\{\Psi_i^{\ell}\}_{\ell\preceq{m_i}});\Psi_{-i}^m|\{\Psi_{-i}^{m'}\}_{m'\prec m}).$$ 
\end{definition}

$\max_{f_m:\Pi_{\ell\preceq m_i}\Sigma_{\ell}\mapsto\Sigma_m}$ means agent $i$ optimize her expected information score over all report strategies that maps her received signals $(\{\Psi_i^{\ell}\}_{\ell\preceq{m_i}})$ to her reported signal for method $m$.


Like we did in the analysis of HMIP, we need to guarantee that for agent $i$ whose performed method is $m_i$, the amount of her received information defined by the above definition should be her optimal payment in Multi-HMIM, given that Multi-HMIM has access to all levels of honest signals reported by other agents. Note that the building block $Corr$ in our mechanism is an unbiased estimator of $MI^{tvd}$ only if the signals are positively correlated. Thus, in order to make the above guarantee, we make an additional assumption---positively correlated guess: agents' optimal guesses for each method $m$'s output are positively correlated with $m$'s real output.

\begin{assumption}[Positively correlated guess]~\label{assume:pcg}
For agent $i$ whose performed method is $m_i$, for all $m$, for all subset $M'\subset \{m'|m'\prec m\}$, there exists $f_{m,M'}^*$ such that
$$f_{m,M'}^*\in \argmax_{f_m:\Pi_{\ell\preceq m_i}\Sigma_{\ell}\mapsto\Sigma_m} MI^{tvd}(f_m(\{\Psi_i^{\ell}\}_{\ell\preceq{m_i}});\Psi_{-i}^m|\{\Psi_{-i}^{m'}\}_{m'\in M'})$$ and $f_{m,M'}^*(\{\Psi_i^{\ell}\}_{\ell\preceq{m_i}})$ is positively correlated with $\Psi_{-i}^m$.
\end{assumption}

\begin{definition}[Prudent strategy in Multi-HMIM]
For each agent $i$, we say she plays prudent strategy in Multi-HMIM($\{\alpha_m\}_m$) if she (a) performs method $m_i^*$ for all her tasks such that 
\begin{align*}
    m_i^*=\argmax_{m_i}\left(AOI(m_i,\text{Multi-HMIM}(\{\alpha_m\}_m))-h_i(m_i)\right);
\end{align*} (b) reports her method $m_i^*$ honestly and reports her all received signals honestly for all her tasks.  
\end{definition}


\begin{definition}[\Faithful  coefficients for Multi-HMIM]\label{def:prudentcoefficients}
Given the priors $\{Q_m\}_m$, we say the coefficients $\{\alpha_m\}_m$ are \faithful for Multi-HMIM($\{\alpha_m\}_m$)  if given the coefficients $\{\alpha_m\}_m$, for every maximal $m$, for every task $t$, \emph{among the agents who are assigned task $t$}, there exists at least \textbf{two} agents whose prudent strategy in Multi-HMIM($\{\alpha_m\}_m$)  are performing method $m$. 
\end{definition}


\begin{definition}[Truthful strategy in Multi-HMIM] For each agent $i$, we say she plays truthful strategy if for each task $t$, she honestly report her method $m_i(t)$ for task $t$ and for each $m\prec m_i(t)$, either she chooses to not report or she reports honestly. \end{definition}

We allow agents to guess the signals they did not receive. Thus, in the definition of prudent strategy and truthful strategy, we only require agents to honestly report the signals they receive and do not put any restriction on their guesses. 

Here we propose a new mechanism design goal: we say a mechanism is (strictly) \emph{truthful} if for each agent, when she believes other agents play a truthful strategy, she can (strictly) maximize her expected utility by playing a truthful strategy. 

The truthful property is incomparable with the \faithful property. A \faithful mechanism incentivizes the efforts of agents but it requires agents to believe other agents play prudent strategy. A truthful mechanism may not be able to incentivize efforts of agents but it incentivizes truthful report by only requiring agents to believe other agents either report honestly or choose to not report.

\begin{theorem}\label{thm:multi}
With Assumption~\ref{assume:a priori}, \ref{assume:pc}, \ref{assume:id}, Multi-HMIM($\{\alpha_m\}_m$) is truthful; moreover, when $\{\alpha_m\}_m$ are \faithful for Multi-HMIM($\{\alpha_m\}_m$), Multi-HMIM($\{\alpha_m\}_m$) is \faithful and truthful. \end{theorem}


In order to show the truth property of Multi-HMIM, we will show for each agent, given that other agents play truthful strategy, (1) conditioning on using pure effort strategy, she can maximize her payment as well as her utility by reporting all her received information honestly; (2) pure effort strategy gives her better utility than mixed effort strategy. We can apply Theorem~\ref{thm:HMIP} directly and use the information monotonicity of $MI^{tvd}$ to prove part (1) directly. In order to show part (2), we need to solve the \emph{mixed effort strategy problem} in the multi-task setting---agents put high level effort only for a partial number of tasks but claim that they spend high level effort all the time. Note that even though agents can expend lower effort in randomizing between performing a low level method and a high level method than purely performing high level method, they also obtain lower payment since they have less ``agreement'' with high level information provided by other people. It turns out that the convexity of the $f$-mutual information---including $MI^{tvd}$---implies that agents cannot obtain higher \emph{utility}---which is the payment minus the cost---by playing a mixed effort strategy. The \faithful property immediately follows from the truthful property and the condition that the coefficients are \faithful. We defer the formal proof to the appendix.


\else

With the positively correlated and other additional natural assumptions (see full version), with similar definitions for the amount of information and \faithful coefficients with HMIP, we can show Multi-HMIM is \faithful as well by mapping it into the HMIP framework. Moreover, we define truthful strategy in Multi-HMIM as the strategy where the agent honestly reports her performed method and that method's output for each task; for other methods' outputs, either she chooses to not report or she reports honestly. We can show Multi-HMIM is truthful: for each agent, when she believes other agents play a truthful strategy, she can maximize her expected utility by playing a truthful strategy as well. We defer the formal analysis of Multi-HMIM to the full version.

\fi

\setlength{\textfloatsep}{0pt}
\begin{algorithm}[t]
\caption{Building Block $Corr$}\label{corr}
\begin{algorithmic}[1]
\Procedure{$Corr$}{$\bm{v}_1;\bm{v}_2$} \Comment{e.g. $\bm{v}_1=(\smiley, \emptyset,\smiley,\smiley,\smiley)$, $\bm{v}_2=(\smiley,\smiley,\smiley,\smiley,\emptyset)$}
\If {either $\bm{v}_1$ or $\bm{v}_2$ has fewer than two non-empty entries}  \Return {0} 
\Else 
\State{$B\subset [M] \gets$  the set of entries where both $\bm{v}_1$ and $\bm{v}_2$ are not empty } 
\Comment{\emph{$B\gets\{1,3,4\}$}}
\If {$B=\emptyset$} \Return {0} 
\Else 
\For {$t_B\in B$} \Comment{\emph{We call $t_B$  a \textbf{reward task}}}
\State {$v_1(t_1)\gets$ a random non-empty entry in $\bm{v}_1$} \\ \Comment{\emph{$v_1(t_1)\gets\smiley$}}
\State {$v_2(t_2)\gets$ a random non-empty entry in $\bm{v}_2$, $t_2\neq t_1$} \Comment{\emph{$v_2(t_2)\gets\smiley$}}
\State {$Corr_{t_B}\gets \mathbbm{1}(v_1(t_B)=v_2(t_B))-\mathbbm{1}(v_1(t_1)=v_2(t_2))$}  \Comment{\emph{$Corr_{t_B}\gets 0$}}
\EndFor 
\State{\Return {$\sum_{t_B\in B} Corr_{t_B}$ and ``success''}} \Comment{\emph{Return 0}}
\EndIf
\EndIf
\EndProcedure

\noindent\rule{12.5cm}{0.01pt}

\Procedure{$Corr$}{$\bm{v}_1;\bm{v}_2|V$} \Comment{e.g. $\bm{v}_1=(\smiley, \smiley,\frownie,\smiley,\smiley)$, $\bm{v}_2=(\smiley,\smiley,\frownie,\smiley,\frownie)$, $V=\{v\}$, $v=(\smiley, \smiley,\frownie,\smiley,\frownie)$}
\State {$C\gets $ the set of entries where every $v\in V$ is not empty} \Comment{$C\gets \{1,2,3,4,5\}$}
\If {$C=\emptyset$} \Return {$Corr(\bm{v}_1;\bm{v}_2)$}
\Else 
\State {$t_C^*\gets$ a random element in $C$} \Comment{$t_C^*\gets 2$}
\State {$D\gets \emptyset$}
\For {$t\in [T]$} 
\If{for every $v\in V$, $v(t)=v(t_C^*)$}
\State {put $t$ in $D$} \EndIf \EndFor \Comment{$D= \{1,2,4\}$, $\bm{v}_1(D)=\bm{v}_2(D)=(\smiley,\smiley,\smiley)$}

\State{\Return{$Corr(\bm{v}_1(D);\bm{v}_2(D))$}} \\ \Comment{Return $Corr(\bm{v}_1(D);\bm{v}_2(D))=0$ and ``success''}
\EndIf
\EndProcedure
\end{algorithmic}
\end{algorithm}

\subsection{Learning Information Structure with a Large Number Tasks}\label{subsec:multi}

%
\begin{assumption}[$\delta_0$-gap]\label{assume:gap}
For each $m$, we assume that for every $i\neq j$, each $m'\neq m$
$$MI^f(\Psi_i^m;\Psi_j^m)>\frac{1}{\delta_0}\qquad MI^f(\Psi_i^m;\Psi_j^{m'})<\frac{1}{\delta_0}$$
\end{assumption}

The above assumption guarantees that when we can accurately learn the $f$-mutual information between two agents' answer vectors, we can accurately classify the answer vectors and then learn the maximal method's outputs correctly.

\paragraph{Learning-based Multi-HMIM($\mathcal{L}$)}\begin{description}
\item [Report] Each agent $i$ is assigned $T$ tasks and asked to perform the same method for all tasks. For agent $i$ who performs method $m_i$, she is asked to report her own answer vector $$\bm{{\psi}}_i^{m_i}=(\psi_{i}^{m_i}(1),\psi_{i}^{m_i}(2),...,\psi_{i}^{m_i}(T))$$ and, for each method $m\neq m_i$, is asked to optionally report her answer vector $\bm{{\psi}}_i^m$. We denote the set of methods whose outputs are reported by agent $i$ as $\methodset_i$ and the actual answer vector she reports for each method $\ell\in\methodset_i$ as $\bm{\hat{\psi}}_i^{\ell}$. Agent $i$ can name the methods freely\footnote{The mechanism will ignore the name of the methods and only record the relationship that the other answer vectors reported by agent $i$ have lower level than agent $i$'s own answer vector.}.





\item [Learning Information Structure]


We define the distance between $\bm{\hat{\psi}}_i^m$ and $\bm{\hat{\psi}}_j^{m'}$ as $\frac{1}{MI^{f}(\hat{\Psi}_i^m;\hat{\Psi}_j^{m'})}$. The mechanism starts to cluster answer vectors. A set of answer vectors are clustered into one cluster if and only if their pairwise distance is less than $\delta_0$. A cluster may have $\geq 1$ answer vector(s). 

For two clusters $m_1,m_2$, $m_1\succ m_2$ if and only if there exists an agent who's own answer vector is in cluster $m_1$ and also provides an answer vector which is classified in cluster $m_2$. The mechanism picks positive real values for the type payment scale  $\alpha_m$ according to a rule $\mathcal{L}$.

\item [Information Score] The mechanism learns the information structure using all agents' reports excluding agent $i$. We denote the set of clusters by $\methodset_{-i}$. For each cluster $m\in \methodset_{-i}$, the mechanism randomly picks an answer vector, denoted $\bm{\hat{\psi}}_{-i}^m$, from it.

Agent $i$ is paid her information score: $$\sum_{m\in\methodset_{-i}}\alpha_m MI^{f}(\{\hat{\Psi}_i^{\ell}\}_{\ell\in\methodset_i};\hat{\Psi}_{-i}^m|\{\hat{\Psi}_{-i}^{m'}\}_{m'\prec m,m'\in \methodset_{-i}})$$

which can be calculated accurately when the number of tasks is large. 
\end{description}

We define $\bm{\alpha}(\mathcal{L}):=\{\alpha_m(\mathcal{L})\}_m$ as the coefficients determined by $\mathcal{L}$, given that the mechanism has access to all levels of honest answer vectors. Here the amount of information and prudent strategy are defined by the same way in HMIP, except that the coefficients are $\bm{\alpha}(\mathcal{L})$.  

\ifnum\fullversion = 1

\begin{definition}[Prudent strategy in Learning-based Multi-HMIM]\label{def:prudentstrategy}
For each agent $i$, we say she plays a prudent strategy in Learning-based Multi-HMIM($\mathcal{L}$) if she chooses to (a) perform method $m_i^*$ such that 
$$m_i^*\in \argmax_{m_i}\left( AOI(m_i,\text{HMIP}(MI^f,\{\alpha_m(\mathcal{L})\}_m)) -h_i(m_i)\right);$$ (b) report all received information honestly. 
\end{definition}


We also define \faithful $\mathcal{L}$ such that $\bm{\alpha}(\mathcal{L})$ is \faithful in the definition in HMIP. 

\begin{definition}[\Faithful rule for Learning-based Multi-HMIM]\label{def:prudent rule}
Given the priors $\{Q_m\}_m$, we say the rule $\mathcal{L}$ that determines the coefficients is \faithful for Learning-based Multi-HMIM($\mathcal{L}$)  if given $\mathcal{L}$, for every maximal $m$, there exists at least \textbf{two} agents whose prudent strategy in Learning-based Multi-HMIM($\mathcal{L}$)  are performing method $m$. 
\end{definition}

\begin{theorem}\label{thm:learning}

With Assumption~\ref{assume:a priori}, Learning-based multi-HMIM is dominant truthful.

Moreover, with Assumption~\ref{assume:gap}, when the rule $\mathcal{L}$ is \faithful, Learning-based multi-HMIM is \faithful, dominant truthful and will output the hierarchical information structure as well as the maximal level(s) answer vector given that agents play prudent strategy. \end{theorem}

\else
We also define \faithful $\mathcal{L}$ such that $\bm{\alpha}(\mathcal{L})$ is \faithful in the definition in HMIP. We can show with the a priori similar and random order assumption, Learning-based multi-HMIM is dominant truthful. With the gap assumption, when the rule $\mathcal{L}$ is \faithful, Learning-based multi-HMIM is \faithful, dominant truthful and will output the hierarchical information structure as well as the maximal level(s) answer vector given that agents play prudent strategy.

\fi

Learning-based multi-HMIM can be mapped to HMIP since a large number of tasks and the gap assumption allow the estimation of the prior and the learning of the information structure and correct clusters. Note that even if the mechanism clusters incorrectly, the mechanism is still dominant truthful since even each agent is paid by the mutual information between her information and ``wrong'' information, the information monotonicity still incentivizes the agent to report all information she has. Thus we do not need the gap assumption for the dominant truthfulness. With the gap assumption, we can cluster agents correctly and use Theorem~\ref{thm:HMIP} to show the \faithful property. \ifnum\fullversion=1 We defer the formal proof to the appendix. \else We defer the formal statement and proof to the full version. \fi  Moreover, we want to emphasize that our mechanisms work even if every agent only has a piece of correct information for the information structure.

\section{Single-task Setting}\label{sec:single}
In the single task setting without known prior, the literature usually assumes a common prior assumption and follows a signal-prediction framework \cite{prelec2004bayesian}---asking agents not only her signal but also her prediction. To achieve truthfulness for $\geq 3$ agents, \citet{radanovic2014incentives} and \citet{KG16} punish each agent if her predictions differ from the predictions of other agents who report the same signals with her, and reward each agent for the accuracy of her prediction. Therefore, for each agent, to maximize her accuracy reward, she will honestly report her predictions. To avoid the punishment for the ``inconsistency'', she will honestly report her received signals as well because of the following commonly assumed assumption\ifnum\fullversion=0 ---\emph{common prior and stochastic relevance}: for every agent $i$ and $j$, they will have the same belief for the distribution of the signals received by other agents if and only if they receive the same signals. \else.


\begin{assumption}[common prior and stochastic relevance]\label{assume:cp}
We assume that for every agent $i$ and $j$, they will have the same belief for the distribution of the signals received by other agents if and only if they receive the same signals. 
\end{assumption}

\fi


\subsection{Applying HMIP in the Single-task Setting}

We follow the previous ``signal-prediction'' framework and ``punishing inconsistency'' idea in the hierarchical information case. We ask agents their received signals and predictions for different levels. We pay each agent the accuracy of her forecasts. The high expertise agents have accurate predictions for even high cost information reports while the low expertise agents only have accurate prediction for low cost information. Therefore, high expertise agents will be paid more.


\paragraph{Single-HMIM$(PS,\{\alpha_m\}_m)$}\begin{description} 
\item [Report (signals, predictions)] Each agent $i$ who performs method $m_i$ is asked to report her received signals $\{\sigma_i^m\}_{m\preceq m_i}$ and her forecast $p_{i}^{m_i}$ for $\Psi_{-i}^{m_i}$. For each $m\neq m_i$, she is asked to optionally report her forecast $p_{i}^{m}$ for $\Psi_{-i}^m$. We denote her report for her received signals as $\{\hat{\sigma}_i^m\}_{m\preceq\hat{m}_i}$ and her prediction report as $\{\hat{p}_i^m\}_{m \in \methodset_i}$ where $\methodset_i$ is the set of methods whose outputs are predicted by agent $i$. 

\item [Prediction Score] We define $\methodset_{-i}$ as the set of methods whose outputs are reported by an agent who is not agent $i$. For each $m \in\methodset_{-i}$, we pick an arbitrary reference agent $j\neq i$ whose performed method is higher than $m$ and denote his report for method $m$'s output by $\hat{\sigma}^m$. Agent $i$'s prediction score is $\sum_{m \in \methodset_{-i}\cap \methodset_{i}} \alpha_m PS(\sigma^m,\hat{p}_i^m)$.

\item [Information Score] If there is no other agent who reports the same signals as agent $i$, then agent $i$'s information score is $0$. Otherwise, arbitrarily pick a reference agent $j\neq i$ from the agents who report the same signals as agent $i$. Agent $i$'s information score is minus the inconsistency between her prediction report and agent $j$'s prediction report, that is, $$-\left(\sum_{m \in\methodset_i\cap \methodset_j}\alpha_m(PS(\hat{p}_j^m,\hat{p}_j^m)-PS(\hat{p}_j^m,\hat{p}_i^m))\right).$$ 
\end{description}

In Single-HMIM, the payment of each agent is $\alpha* \textit{Information Score}+\beta* \textit{Prediction Score}.$


\begin{definition}[Truthful strategy in Single-HMIM] For each agent $i$ who performed method $m_i$, she plays truthful strategy if she honestly report her received signals $\{\sigma_i^m\}_{m\preceq m_i}$ and forecast for $\Psi_{-i}^{m_i}$ and for each $m\neq m_i$, either she does not report or she reports her forecast for $\Psi_{-i}^m$ honestly. 
\end{definition}

\ifnum\fullversion=1

We denote $p_{m_i}^m$ as agent $i$'s honest forecast for $\Psi_{-i}^m$ given that she performs method $m_i$.

\begin{definition}[Amount of information in Single-HMIM]
For each agent $i$ who performs method $m_i$, her acquired amount of information is defined as $$AOI(m_i,\text{Single-HMIM}(PS,\{\alpha_m\}_m)):=\sum_{m}\alpha_m \E_{Q_m}[PS(\sigma^m,p_{m_i}^{m})].$$
\end{definition}

Later in the proof of Theorem~\ref{thm:single}, we will see the amount of information is also the optimal payment of agent $i$ who performs method $m_i$ in Single-HMIM, given that Single-HMIM has access to all levels of honest signals reported by other agents.

\begin{definition}[Prudent strategy in Single-HMIM]\label{def:prudentstrategy}
For each agent $i$, we say she plays a prudent strategy in Single-HMIM$(PS,\{\alpha_m\}_m)$ if she chooses to (a) perform method $m_i^*$ such that 
$$m_i^*\in \argmax_{m_i}\left( AOI(m_i,\text{Single-HMIM}(PS,\{\alpha_m\}_m))-h_i(m_i)\right);$$ (b) play a truthful strategy. 
\end{definition}

\begin{definition}[\Faithful coefficients for Single-HMIM]\label{def:prudentcoefficients}
Given the priors $\{Q_m\}_m$, we say the coefficients $\{\alpha_m\}_m$ are \faithful for Single-HMIM($PS$,$\{\alpha_m\}_m$) if given the coefficients $\{\alpha_m\}_m$, for every maximal $m$, there exists at least \textbf{two} agents whose prudent strategy in Single-HMIM($PS$,$\{\alpha_m\}_m$) is performing method $m$. 
\end{definition}

Recall that a mechanism is (strictly) \emph{truthful} if for each agent when she believes other agents play a truthful strategy, she can (strictly) maximize her expected utility by playing a truthful strategy.

\begin{theorem}\label{thm:single}
With Assumption~\ref{assume:cp}, single-HMIM is strictly truthful; moreover, when the coefficients is \faithful for single-HMIM, single-HMIM is \faithful and strictly truthful. 
\end{theorem}

\else

With proper definition for the amount of information and \faithful coefficients, we will show with the common prior assumption, single-HMIM is strictly truthful; moreover, when the coefficients are \faithful for single-HMIM, single-HMIM is \faithful and strictly truthful.

\fi

The strictly truthful property follows from the common prior and stochastic relevance assumption. The \faithful property follows from the definition of prudent strategy and \faithful coefficients. We defer the formal proof to \ifnum\fullversion=1 appendix\else the full version\fi.



\section{Discussion and Future Work}\label{sec:discussion}

Although our work is theoretical, it is not far from applications. For example, in some situations, we could simplify the information structure by roughly dividing it into two levels where the higher level requires more time. Then we can apply our Multi-HMIM or Single-HMIM to peer grading or any other situations where certain agents are only given 15 seconds to grade a work and others are expected to do a good job. We value the information \emph{conditioning on} the reports provided by the ``15 seconds'' agents. We could also use machine learning to obtain the lower level information.

Race, sex, and other stereotypes can be used as  ``cheap" signals.  One  possible future application of this work is to address fairness in crowdsourcing.  

There are several limitations of our mechanisms.  
Our Multi-HMIM requires that the hierarchy structure of the information is common knowledge.  Our key assumption, that more sophisticated agents know the information of less sophisticated agents may not always hold, especially if the less sophisticated agents collude to report seemingly irrelevant information (e.g.,\ the third letter of the third word).  

Our analysis of the Learning-based HMIM requires the agents to perform an infinite number of tasks.  One future direction is having the sample complexity analysis.   Our Single-HMIM still requires a forecast report.  A future direction is making the forecast report optional  in some scenarios (we might ask agents for the most common answer if they believe it is not their reported answer). Another future direction is the information cost elicitation: tuning the coefficients of the mechanisms such that the payment matches the actual effort required by agents.

A key future direction is to test our mechanism and explore the limitations in practice by performing real-world experiments.  This would require designing a suitable user interface.

\bibliographystyle{ACM-Reference-Format}
\bibliography{refs}

\ifnum\fullversion = 1

\newpage

\appendix

\section{Additional proofs}
{
\renewcommand{\thetheorem}{\ref{lem:psim}}
\begin{fact}[Information monotonicity of proper scoring rules]
Given any strictly proper scoring rule $PS$,  
$$\E_{X,Y,Z} PS(Y,\Pr[\bm{Y}|X,Z])\geq \E_{X,Y} PS(Y,\Pr[\bm{Y}|X]).$$
The equality holds if and only if $\Pr[\bm{Y}|X=x,Z=z]=\Pr[\bm{Y}|X=x]$ for all $(x,z)$ where $\Pr[X=x,Z=z]>0$.
\end{fact}
\addtocounter{theorem}{-1}
}

\begin{proof}
\begin{align*}
    \E_{X,Y} PS(Y,\Pr[\bm{Y}|X])&= \sum_{x,y}\Pr[X=x,Y=y] PS(Y=y,\Pr[\bm{Y}|X=x])\\
    &= \sum_{x,y,z}\Pr[X=x,Y=y,Z=z] PS(Y=y,\Pr[\bm{Y}|X=x])\\
    &= \sum_{x,z}\Pr[X=x,Z=z]\sum_y\Pr[Y=y|X=x,Z=z] PS(Y=y,\Pr[\bm{Y}|X=x])\\ 
      &= \sum_{x,z}\Pr[X=x,Z=z]PS(\Pr[\bm{Y}|X=x,Z=z],\Pr[\bm{Y}|X=x])\\ \tag{$PS$ is strictly proper}
    &\leq \sum_{x,z}\Pr[X=x,Z=z]PS(\Pr[\bm{Y}|X=x,Z=z],\Pr[\bm{Y}|X=x,Z=z])\\
    &=\E_{X,Y,Z} PS(Y,\Pr[\bm{Y}|X,Z])
\end{align*}
The equality holds if and only if $\Pr[\bm{Y}|X=x,Z=z]=\Pr[\bm{Y}|X=x]$ for all $(x,z)$ where $\Pr[X=x,Z=z]>0$ since $PS$ is \emph{striclty} proper.

\end{proof}

{
\renewcommand{\thetheorem}{\ref{thm:multi}}

\begin{theorem}
With Assumption~\ref{assume:a priori}, \ref{assume:pc}, \ref{assume:id}, Multi-HMIM($\{\alpha_m\}_m$) is truthful; moreover, when $\{\alpha_m\}_m$ are \faithful for Multi-HMIM($\{\alpha_m\}_m$), Multi-HMIM($\{\alpha_m\}_m$) is \faithful and truthful. \end{theorem}

\addtocounter{theorem}{-1}
}

\begin{proof}

Since we assume all tasks are a priori similar, without loss of generality, we can assume every agent uses the same (possibly mixed) report and (possibly mixed) effort strategy for all tasks.


\paragraph{Truthful} We divide the proof into two parts. For each agent $i$, given that she believes other agents report honestly (may not report all signals they have), we will show (1) conditioning on agent $i$ playing pure effort strategy, she should maximize her payment as well as the utility by playing truthful strategy; (2) it's better for agent $i$ to play pure effort strategy---performing the same method all the time---than mixed effort strategy.

Part (1). We want to show that for each agent $i$ who always perform method $m_i$, given other agents honestly report their methods and signals, for each $m\preceq m_i$, she should honestly report her real signal $\psi_i^m$ to maximize her expected information score in $m$'s level, that is,  $$\E[2 \alpha_m Corr(\bm{\hat{\psi}}_i^m;\bm{\hat{\psi}}_{-i}^m|\bm{\{\hat{\psi}}_{-i}^{m'}\}_{m'\prec m}].$$ Since we assume other agents report honestly and we have assumed that the signals agents receive for every method are homogeneous, we replace $\bm{\hat{\psi}}_{-i}^m,\bm{\hat{\psi}}_{-i}^{m'}$ by $\bm{{\psi}}_{-i}^m,\bm{{\psi}}_{-i}^{m'}$.  

When we run algorithm~\ref{corr} to calculate $Corr(\bm{\hat{\psi}}_i^m;\bm{\hat{\psi}}_{-i}^m|\bm{\{\hat{\psi}}_{-i}^{m'}\}_{m'\prec m})$, in the situation the algorithm does not return ``success''---situation 0---her information score in $m$'s level is 0 regardless of agent $i$ reports for method $m$'s output. In the situation the algorithm returns ``success'', either it runs $Corr(\bm{\hat{\psi}}_i^m;\bm{\hat{\psi}}_{-i}^m)$ and returns ``success''---situation 1---or it runs $Corr(\bm{\hat{\psi}}_i^m(D);\bm{\hat{\psi}}_{-i}^m(D))$ and returns ``success''---situation 2.

For each task, each $m$, fixing agents' choices for whether to provide a signal or $\emptyset$, the situation which the algorithm runs in is fixed as well. We only need to consider each situation separately.

\begin{claim}\label{claim:1}
Given that other agents report honestly, for each agent $i$ who always perform $m_i$, for all $m\preceq m_i$, when agent $i$ honestly reports method $m$'s output, her expected information score in $m$'s level per each reward task is 
 
 $$  \alpha_m MI^{tvd}({\Psi}_i^m;{\Psi}_{-i}^m)$$
 in situation 1;
 
 $$  \alpha_m MI^{tvd}({\Psi}_i^m;{\Psi}_{-i}^m|\{{\Psi}_{-i}^{m'}\}_{m'\prec m})$$
 in situation 2.
\end{claim}

\begin{claim}\label{claim:2}
Given that other agents report honestly, for each agent $i$, when agent $i$ reports method $m$'s output as $\hat{\psi}_i^m$, her expected information score in $m$'s level per each reward task is $\leq$ 
 $$ \alpha_m MI^{tvd}(\hat{\Psi}_i^m;{\Psi}_{-i}^m)$$ in situation 1;
 
 $$  \alpha_m MI^{tvd}(\hat{\Psi}_i^m;{\Psi}_{-i}^m|\{{\Psi}_{-i}^{m'}\}_{m'\prec m})$$ in situation 2. The equality holds if $\hat{\Psi}_i^m$ is positively correlated with ${\Psi}_{-i}^m$ (conditioning on $\{{\Psi}_{-i}^{m'}\}_{m'\prec m}$). 
\end{claim}

  Once we show the above two claims. Since
 
 \begin{align*}
     & MI^{tvd}(\hat{\Psi}_i^m;{\Psi}_{-i}^m|\{{\Psi}_{-i}^{m'}\}_{m'\prec m})\\ \tag{Agent $i$ uses the report strategy $f_m$ to report $m$'s output}
     &= MI^{tvd}(f_m(\{\Psi_i^{m'}\}_{m'\preceq  m_i});{\Psi}_{-i}^m|\{{\Psi}_{-i}^{m'}\}_{m'\prec m})\\\tag{Information Monotonicity of $MI^f$}
     &\leq MI^{tvd}(\{\Psi_i^{m'}\}_{m'\preceq  m_i};{\Psi}_{-i}^m|\{{\Psi}_{-i}^{m'}\}_{m'\prec m})\\ \tag{Assumption~\ref{assume:id}}
     &=MI^{tvd}({\Psi}_i^m;{\Psi}_{-i}^m|\{{\Psi}_{-i}^{m'}\}_{m'\prec m})
 \end{align*}
 
 and similarly $MI^{tvd}(\hat{\Psi}_i^m;{\Psi}_{-i}^m)\leq MI^{tvd}({\Psi}_i^m;{\Psi}_{-i}^m)$. Part (1) follows immediately.

Part (2). This part is implied by the complexity of $MI^{tvd}$. We give a formal proof here. We consider situation 1 here. For any $0\leq \lambda\leq 1$, any two methods $m_1,m_2$, if agent $i$ perform method $m_1$ with probability $\lambda$, method $m_2$ with probability $1-\lambda$, for every $m$, agent $i$'s utility in $m$'s level is less than

\begin{align*}
    & MI^{tvd}(\hat{\Psi}_i^m;{\Psi}_{-i}^m)-(\lambda h_i(m_1)+(1-\lambda) h_i(m_2))\\ 
&\leq \max_{f_m} MI^{tvd}(f_m(\text{her received signals)};{\Psi}_{-i}^m)-(\lambda h_i(m_1)+(1-\lambda) h_i(m_2))\\ \tag{$f_m^*$ is the optimal report strategy.}
&=MI^{tvd}(f_m^*(\text{her received signals)};{\Psi}_{-i}^m)-(\lambda h_i(m_1)+(1-\lambda) h_i(m_2))\\ \tag{Convexity of $MI^f$}
&\leq\lambda (MI^{tvd}(f_m^*(\{\Psi_i^{m'}\}_{m'\preceq  m_1});{\Psi}_{-i}^m)-h_i(m_1))+ (1-\lambda) (MI^{tvd}(f_m^*(\{\Psi_i^{m'}\}_{m'\preceq  m_2});{\Psi}_{-i}^m)-h_i(m_2))\\
&\leq \max\{MI^{tvd}(f_m^*(\{\Psi_i^{m'}\}_{m'\preceq  m_1});{\Psi}_{-i}^m)-h_i(m_1),MI^{tvd}(f_m^*(\{\Psi_i^{m'}\}_{m'\preceq  m_2});{\Psi}_{-i}^m)-h_i(m_2)\}
\end{align*}

in situation 1. Without loss of generality, we assume $$MI^{tvd}(f_m^*(\{\Psi_i^{m'}\}_{m'\preceq  m_1});{\Psi}_{-i}^m)-h_i(m_1)\geq MI^{tvd}(f_m^*(\{\Psi_i^{m'}\}_{m'\preceq  m_2});{\Psi}_{-i}^m)-h_i(m_2).$$

Then 
\begin{align*}
    & MI^{tvd}(\hat{\Psi}_i^m;{\Psi}_{-i}^m)-(\lambda h_i(m_1)+(1-\lambda) h_i(m_2))\\
&\leq MI^{tvd}(f_m^*(\{\Psi_i^{m'}\}_{m'\preceq  m_1});{\Psi}_{-i}^m)-h_i(m_1)\\
& \leq \max_{f_m} MI^{tvd}(f_m(\{\Psi_i^{m'}\}_{m'\preceq  m_1});{\Psi}_{-i}^m)-h_i(m_1)
\end{align*} 
The analysis for situation 2 is similar. With the positively correlated guess assumption (Assumption~\ref{assume:pcg}) and Claim~\ref{claim:2}, we know $\max_{f_m} MI^{tvd}(f_m(\{\Psi_i^{m'}\}_{m'\preceq  m_1});{\Psi}_{-i}^m)$ can be obtained by agent $i$ in Multi-HMIM by always performing $m_1$ and playing a proper report strategy. Thus, agent $i$ cannot obtain better utility by playing mixed effort strategy.

\paragraph{\Faithful} We can follow the proof of truthful property and additionally show that when the coefficients are \faithful, for each agent $i$, when she believes others agents play prudent strategy, agent $i$ should pick the effort strategy defined by the prudent strategy as her optimal effort strategy. When the coefficients are \faithful, based on the definition of \faithful coefficients, for each agent $i$, when she believe other agents play prudent strategy, for each task she finished, there must exists another agent who finished the same task with her, using the method that is higher or equal to her. Thus, agent $i$'s all tasks are reward tasks for her, and algorithm~\ref{corr} will always run into situation 2 since the mechanism always has access to all levels of information. With the positively correlated guess assumption (Assumption~\ref{assume:pcg}) and Claim~\ref{claim:2}, agent $i$'s optimal utility is proportional to 

$$\sum_{m\in M}\max_{f_m:\Pi_{\ell\preceq m_i}\Sigma_{\ell}\mapsto\Sigma_m}\alpha_m MI^{tvd}(f_m(\{\Psi_i^{\ell}\}_{\ell\preceq{m_i}});\Psi_{-i}^m|\{\Psi_{-i}^{m'}\}_{m'\prec m}) -h_i(m_i)$$ by always performing method $m_i$. Thus, agent $i$'s optimal effort strategy should be the effort strategy defined by the prudent strategy, given that she believes other agents play prudent strategy. 
\end{proof}

 {
\renewcommand{\thetheorem}{\ref{thm:learning}}

 \begin{theorem}

With Assumption~\ref{assume:a priori}, Learning based multi-HMIM is dominant truthful.

Moreover, with Assumption~\ref{assume:gap}, when the rule $\mathcal{L}$ is \faithful, Learning based multi-HMIM is \faithful, dominant truthful and will output the hierarchical information structure as well as the maximal level(s) answer vector given that agents play prudent strategy. \end{theorem}
\addtocounter{theorem}{-1}
}

\begin{proof}[Proof for Theorem~\ref{thm:learning}]
Since we assume all tasks are a priori similar, without loss of generality, we can assume every agent use the same report and effort strategy for all tasks.

In order to show the dominant truthful property, we will show for each agent, fixing any other agents' strategies, (1) conditioning on using pure effort strategy, she can maximize her payment as well as the utility by reporting her received information honestly; (2) pure effort strategy has higher utility than mixed effort strategy. 

Part (1). Even if the mechanism clusters incorrectly, part (1) still follows directly from the information monotonicity property of $f$-mutual information $MI^{f}$. 

Part (2). The proof here is the same with the part (2) proof in Theorem~\ref{thm:multi}. We give a formal proof here. 

For any $0\leq \lambda\leq 1$, any two methods $m_1,m_2$, if agent $i$ perform method $m_1$ with probability $\lambda$, method $m_2$ with probability $1-\lambda$,  agent $i$'s utility in $m$'s level is

\begin{align*}
    & MI^{f}(\text{her reported signals};\hat{\Psi}_{-i}^m|\{\hat{\Psi}_{-i}^{m'}\}_{m'\prec m,m'\in \methodset_{-i}})-(\lambda h_i(m_1)+(1-\lambda) h_i(m_2))\\ 
\leq& MI^{f}(\text{her received signals};;\hat{\Psi}_{-i}^m|\{\hat{\Psi}_{-i}^{m'}\}_{m'\prec m,m'\in \methodset_{-i}})-(\lambda h_i(m_1)+(1-\lambda) h_i(m_2))\\  \tag{convexity of $MI^f$}
\leq&\lambda (MI^{f}(\{\Psi_i^{m'}\}_{m'\preceq  m_1};\hat{\Psi}_{-i}^m|\{\hat{\Psi}_{-i}^{m'}\}_{m'\prec m,m'\in \methodset_{-i}})-h_i(m_1))\\
&+ (1-\lambda) (MI^{f}(\{\Psi_i^{m'}\}_{m'\preceq  m_2};\hat{\Psi}_{-i}^m|\{\hat{\Psi}_{-i}^{m'}\}_{m'\prec m,m'\in \methodset_{-i}})-h_i(m_2))\\
\leq& \max\{MI^{f}(\{\Psi_i^{m'}\}_{m'\preceq  m_1};\hat{\Psi}_{-i}^m|\{\hat{\Psi}_{-i}^{m'}\}_{m'\prec m,m'\in \methodset_{-i}})-h_i(m_1),\\
&MI^{f}(\{\Psi_i^{m'}\}_{m'\preceq  m_2};\hat{\Psi}_{-i}^m|\{\hat{\Psi}_{-i}^{m'}\}_{m'\prec m,m'\in \methodset_{-i}})-h_i(m_2)\}
\end{align*} Thus, each agent $i$ cannot obtain higher utility by playing a mixed effort strategy.

It remains to show the \faithful property. When the rule is \faithful, for each agent $i$, when she believes other agents play prudent strategy, the mechanism must have access to all levels of honest answer vectors due to the definition of prudent strategy and \faithful rule. With Assumption~\ref{assume:gap}, the mechanism can correctly learn the whole hierarchical information structure without agent $i$'s report and use coefficients $\bm{\alpha}(\mathcal{L})$. Thus, her optimal payment for performing method $m_i$ will be $$\sum_{m\in M}\alpha_m MI^{f}(\{\Psi_i^{\ell}\}_{\ell\preceq{m_i}};\Psi_{-i}^m|\{\Psi_{-i}^{m'}\}_{m'\prec m})$$ due to the information monotonicity of $MI^f$. In this case, her optimal strategy is her prudent strategy. Therefore, learning based Multi-HMIM is \faithful and will output the correct hierarchical information structure as well as the maximal level(s) answer vector(s) when agents play prudent strategy.

\end{proof}

 {
\renewcommand{\thetheorem}{\ref{thm:single}}
\begin{theorem}
With Assumption~\ref{assume:cp}, single-HMIM is strictly truthful; moreover, when the coefficients is \faithful for single-HMIM, single-HMIM is \faithful and strictly truthful. 
\end{theorem}
\addtocounter{theorem}{-1}
}

\begin{proof}[for Theorem~\ref{thm:single}]
For each agent $i$, her highest information score is 0. When she believes all other agents honestly report their signals and predictions, she can obtain her highest prediction score via providing her truthful prediction based on the property of the strictly proper scoring rule. While during the same time, she can obtain 0 (the highest) information score according to the common prior assumption. If agent $i$ tell lies about her predictions, in expectation she will receive strictly lower prediction score since $PS$ is strictly proper. If she honestly provides her predictions but lie for the signals, then she will be punished for her information score with positive probability. Therefore, when agent $i$ believes everyone else tells the truth, honestly reporting her truthful signals and predictions strictly maximize her payment.

It remains to show the \faithful property. In Single-HMIM, when the coefficients are \faithful, for each agent $i$, when she believes other agents play prudent strategy, for each $m$, there must exist a reference agent for agent $i$ who reports method $m$'s output. Thus agent $i$'s optimal expected payment by performing method $m_i$ is $$\sum_{m\in M}\alpha_m \E_{Q_m}[PS(\sigma^m,p_{m_i}^{m})]$$ since her optimal information score is always 0. In this case, agent $i$'s optimal strategy is prudent for her. Therefore, Single-HMIM is \faithful. 
\end{proof}

{
\renewcommand{\thetheorem}{\ref{claim:1}}

\begin{claim}
Given that other agents report honestly, for each agent $i$ who always perform $m_i$, for all $m\preceq m_i$, when agent $i$ honestly reports method $m$'s output, her expected information score in $m$'s level per each reward task is 
 
 $$  \alpha_m MI^{tvd}({\Psi}_i^m;{\Psi}_{-i}^m)$$
 in situation 1;
 
 $$  \alpha_m MI^{tvd}({\Psi}_i^m;{\Psi}_{-i}^m|\{{\Psi}_{-i}^{m'}\}_{m'\prec m})$$
 in situation 2.
\end{claim}
\addtocounter{theorem}{-1}
}

\begin{proof}[Proof for Claim~\ref{claim:1}]
 We first show $$\E[ Corr(\bm{\psi}_i^m;\bm{\Psi}_{-i}^m)]=\frac{1}{2}MI^{tvd}({\Psi}_i^m;{\Psi}_{-i}^m).$$ 
 
 \begin{align*}
      \frac{1}{2}MI^{tvd}({\Psi}_i^m;{\Psi}_{-i}^m) \tag{Definition of $MI^{tvd}$}
     &=\frac{1}{2}\sum_{\sigma,\sigma'} |\Pr[\Psi_i^m=\sigma,\Psi^m_{-i}=\sigma']-\Pr[\Psi_i^m=\sigma]\Pr[\Psi^m_{-i}=\sigma']|\\ 
     &= \frac{1}{2}\sum_{\sigma,\sigma'} \mathbbm{1}(\sigma=\sigma')\left(\Pr[\Psi_i^m=\sigma,\Psi^m_{-i}=\sigma']-\Pr[\Psi_i^m=\sigma]\Pr[\Psi^m_{-i}=\sigma']\right)\\\tag{Assumption~\ref{assume:pc}}
     &+\mathbbm{1}(\sigma\neq\sigma')\left(\Pr[\Psi_i^m=\sigma]\Pr[\Psi^m_{-i}=\sigma']-\Pr[\Psi_i^m=\sigma,\Psi^m_{-i}=\sigma']\right)\\ \tag{Combining like terms, $\Pr[E]-\Pr[\neg E]=2\Pr[E]-1$}
      &=\sum_{\sigma} \left(\Pr[\Psi_i^m=\sigma,\Psi^m_{-i}=\sigma]-\Pr[\Psi_i^m=\sigma]\Pr[\Psi^m_{-i}=\sigma]\right)\\ \tag{see Algorithm~\ref{corr}}
      &=\E[ Corr(\bm{{\psi}}_i^m;\bm{{\psi}}_{-i}^m)]  
 \end{align*}
 
 To show $$\E[ Corr(\bm{{\psi}}_i^m;\bm{{\psi}}_{-i}^m|\{\bm{{\psi}}_{-i}^{m'}\}_{m'\prec m})]=\frac{1}{2}MI^{tvd}({\Psi}_i^m;{\Psi}_{-i}^m|\{{\Psi}_{-i}^{m'}\}_{m'\prec m}),$$ we only need to replace every $\Pr[\cdot]$ in the above equations by $\Pr[\cdot|\{{\Psi}_{-i}^{m'}\}_{m'\prec m}=\{{\sigma}^{m'}\}_{m'\prec m}]$ with putting $\sum_{\{{\sigma}^{m'}\}_{m'\prec m}}\Pr[\{{\Psi}_{-i}^{m'}\}_{m'\prec m}=\{{\sigma}^{m'}\}_{m'\prec m}]$ ahead. Note that assumption~\ref{assume:pc} can be applied to this case as well. 
 \end{proof}
 
 {
\renewcommand{\thetheorem}{\ref{claim:2}}

\begin{claim}
Given that other agents report honestly, for each agent $i$, when agent $i$ reports method $m$'s output as $\hat{\psi}_i^m$, her expected information score in $m$'s level per each reward task is $\leq$ 
 $$ \alpha_m MI^{tvd}(\hat{\Psi}_i^m;{\Psi}_{-i}^m)$$ in situation 1;
 
 $$  \alpha_m MI^{tvd}(\hat{\Psi}_i^m;{\Psi}_{-i}^m|\{{\Psi}_{-i}^{m'}\}_{m'\prec m})$$ in situation 2. The equality holds if $\hat{\Psi}_i^m$ is positively correlated with ${\Psi}_{-i}^m$. 
\end{claim}
\addtocounter{theorem}{-1}
}
 
 \begin{proof}[Proof for Claim~\ref{claim:2}]
 The proof is similar with the proof of Claim~\ref{claim:1}. We only need to replace ${\Psi}_i^m$ by $\hat{\Psi}_i^m$ and change the second equation to greater than, that is, 
 
 \begin{align*}
     &\frac{1}{2}\sum_{\sigma,\sigma'} |\Pr[\Psi_i^m=\sigma,\Psi^m_{-i}=\sigma']-\Pr[\Psi_i^m=\sigma]\Pr[\Psi^m_{-i}=\sigma']|\\  
     &\geq \frac{1}{2}\sum_{\sigma,\sigma'} \mathbbm{1}(\sigma=\sigma')\left(\Pr[\Psi_i^m=\sigma,\Psi^m_{-i}=\sigma']-\Pr[\Psi_i^m=\sigma]\Pr[\Psi^m_{-i}=\sigma']\right)\\\tag{$\sum|x|\geq \sum x $}
     &+\mathbbm{1}(\sigma\neq\sigma')\left(\Pr[\Psi_i^m=\sigma]\Pr[\Psi^m_{-i}=\sigma']-\Pr[\Psi_i^m=\sigma,\Psi^m_{-i}=\sigma']\right).
 \end{align*}
 
 Note that the equality holds if $\hat{\Psi}_i^m$ is positively correlated with ${\Psi}_{-i}^m$. Follow the same proof of Claim~\ref{claim:1}, we finish the proof. 
 \end{proof}
\section{Mutual information calculations}\label{sec:cal}
We show the calculations for the mutual information table.

For the length signal, since agents has no uncertainty for this signal, the mutual information between agent $i$'s length signal and agent $j\neq i$'s length signal will be the entropy of length signal. Recall that we have assumed an essay has long length with probability 0.5. Thus, 
\begin{flalign*}
    MI(length;length) = 0.5*\log(0.5)+0.5*\log(0.5)=0.6931
\end{flalign*}

Since an essay's length is independent with its writing and quality, we have the mutual information between the length signal and writing signal, quality, writing conditioning length, quality conditioning writing and length are all zero.

$\Pr[\Psi_i^{m_w}=\smiley,\Psi_j^{m_w}=\smiley]=0.5*0.9*0.9+0.5*0.1*0.1=0.41$
    
$\Pr[\Psi_i^{m_w}=\smiley,\Psi_j^{m_w}=\frownie]=\Pr[\Psi_i^{m_w}=\frownie;\Psi_j^{m_w}=\smiley]=0.5*0.9*0.1+0.5*0.1*0.9=0.09$

$\Pr[\Psi_i^{m_w}=\frownie,\Psi_j^{m_w}=\frownie]=0.5*0.1*0.1+0.5*0.9*0.9=0.41$

We can put the above joint distribution over $(\Psi_i^{m_w};\Psi_j^{m_w})$ to the formula $MI(X;Y)=\sum_{x,y} \Pr[X=x,Y=y]\log \frac{\Pr[X=x,Y=y]}{\Pr[X=x]\Pr[Y=y]}$ and obtain

\begin{flalign*}
    &MI(length, writing;writing)&\\ 
    =& MI(writing;writing)&\\
    = & MI(\Psi_i^{m_w};\Psi_j^{m_w}) = 0.2218
\end{flalign*}

Note that MI(writing;writing) is not the entropy of the writing signal since it is the mutual information between different agents' writing signals.

Similarly, we can calculate the joint distribution over $(\Psi_i^{m_q},\Psi_i^{m_w},\Psi_j^{m_q},\Psi_j^{m_w})$ and set $\frownie=0$ and $\smiley=1$:

\begin{flalign*}
    &\Pr[\Psi_i^{m_q}=a,\Psi_i^{m_w}=b,\Psi_j^{m_q}=c,\Psi_j^{m_w}=d]\\\tag{when the essay has bad quality, bad writing:}
    =&0.4*0.3^a*0.7^{1-a}*0.1^b*0.9^{1-b}*0.3^c*0.7^{1-c}*0.1^d*0.9^{1-d}&\\ \tag{when the essay has bad quality, good writing:}
    &+0.1*0.3^a*0.7^{1-a}*0.9^b*0.1^{1-b}*0.3^c*0.7^{1-c}*0.9^d*0.1^{1-d}&\\ \tag{when the essay has good quality, bad writing:}
    &+0.1*0.7^a*0.3^{1-a}*0.1^b*0.9^{1-b}*0.7^c*0.3^{1-c}*0.1^d*0.9^{1-d}&\\  \tag{when the essay has good quality, good writing:}
    &+0.4*0.7^a*0.3^{1-a}*0.9^b*0.1^{1-b}*0.7^c*0.3^{1-c}*0.9^d*0.1^{1-d}
\end{flalign*}

The fact that the length signal is independent with writing and quality will ease the calculation a lot since we can ignore the length signal if it only shows in one side when we calculate the mutual information. Moreover, since the length signal has no uncertainty, length|length will be a value without uncertainty and can be ignored in the calculation of mutual information.

Aided by the calculator, we can obtain

\begin{flalign*}
    &MI(length, writing; quality)&\\ 
    =& MI(writing;quality)&\\
    = & MI(\Psi_i^{m_w};\Psi_j^{m_q}) = 0.0185
\end{flalign*}

\begin{flalign*}
    &MI(length, writing; writing|length)&\\ 
    =& MI(writing;writing)= 0.2218;
\end{flalign*}

\begin{flalign*}
    &MI(length, writing, quality;writing)&\\ 
    =& MI(quality,writing;writing)&\\
    = & MI(\Psi_i^{m_w},\Psi_i^{m_q};\Psi_j^{m_w}) = 0.2259
\end{flalign*}

\begin{flalign*}
    &MI(length, writing; quality|writing,length)&\\ 
    =& MI(writing;quality|writing)&\\
    = & MI(writing,quality;writing)-MI(writing;writing)
    = 0.2259 - 0.2218 = 0.0041
\end{flalign*}
    
\begin{flalign*}
    &MI(length, writing, quality;quality)&\\ 
    =& MI(quality,writing;quality)&\\
    = & MI(\Psi_i^{m_w},\Psi_i^{m_q};\Psi_j^{m_q}) = 0.0267
\end{flalign*}

\begin{flalign*}
    &MI(length, writing, quality;writing|length) &\\ 
    =& MI(quality,writing;writing) 
    = 0.2259 
\end{flalign*}

\begin{flalign*}
    &MI(length, writing, quality;quality|writing, length)&\\ 
    =& MI(writing,quality;quality|writing)&\\
    =& MI(writing,quality;quality,writing)-MI(writing,quality;writing)&\\
    = & 0.2374-0.2259 = 0.0115
\end{flalign*}

\begin{flalign*}
    &MI(length, writing, quality;length, writing)&\\ 
    =& MI(length, writing, quality;length)+MI(length, writing, quality;writing|length)&\\
    =& MI(length;length)+MI(length, writing, quality;writing|length)\\
    = & 0.6931+0.2259 = 0.9190
\end{flalign*}

\begin{flalign*}
    &MI(length, writing, quality;length, writing,quality)&\\ 
    =& MI(length, writing, quality;length)+MI(length, writing, quality;writing|length)\\
    &+MI(length, writing, quality;quality|writing, length)&\\
    = & 0.6931+0.2259+0.0115= 0.9305
\end{flalign*}

\else

\fi

\end{document}